\documentclass[sn-mathphys,iicol]{sn-jnl}

\jyear{2022}%

\theoremstyle{thmstyleone}%
\theoremstyle{thmstyletwo}%

\theoremstyle{thmstylethree}%

\usepackage{graphicx}

\usepackage[matrix,arrow,color]{xy}

\DeclareMathOperator*{\timesbig}{\scalerel*{\times}{\textstyle\sum}}
\usepackage{scalerel}

\usepackage{bbm}
\usepackage{bm}

\usepackage{mathrsfs}

\newcommand{\FR}{\mathbbm{R}}     			
\newcommand{\FC}{\mathbbm{C}}     			
\newcommand{\FH}{\mathbbm{H}}     			
\newcommand{\RZ}{\mathbbm{Z}}     			
\newcommand{\PP}{{\mathbbm{P}}}    			

\newcommand{\ident}{\mathbbm{1}}  

\newcommand{\CCC}{\mathscr{C}}

\newcommand{\CH}{\mathcal{H}}

\newcommand{\CI}{\mathcal{I}}

\newcommand{\CCK}{\mathscr{K}}

\newcommand{\CN}{\mathcal{N}}

\newcommand{\CO}{\mathcal{O}}

\newcommand{\CS}{\mathcal{S}}

\newcommand{\CCV}{\mathscr{V}}

\newcommand{\CCX}{\mathscr{X}}

\newcommand{\CCY}{\mathscr{Y}}

\newcommand{\CE}{\mathcal{E}}

\newcommand{\sU}{\mathsf{U}}     			
\newcommand{\sGamma}{\mathsf{\Gamma}}     			

\newcommand{\sG}{\mathsf{G}}
\newcommand{\sT}{\mathsf{T}}

\newcommand{\sL}{\mathsf{L}}

\newcommand{\sH}{\mathsf{H}}
\newcommand{\sSU}{\mathsf{SU}}
\newcommand{\sSL}{\mathsf{SL}}
\newcommand{\sGL}{\mathsf{GL}}

\newcommand{\sSO}{\mathsf{SO}}
\newcommand{\sSpin}{\mathsf{Spin}}

\newcommand{\sfGamma}{{\sf \Gamma}}

\newcommand{\Hilb}{{\sf Hilb}}
\newcommand{\Quot}{{\sf Quot}}
\newcommand{\Ob}{{\sf Ob}}

\newcommand{\Ch}{\mathsf{Ch}}

\newcommand{\ii}{\mathrm{i}}   			
\newcommand{\ee}{\mathrm{e}}   			
\newcommand{\dd}{\mathrm{d}}     			
\newcommand{\Tr}{\mathrm{Tr}}
\newcommand{\ch}{\mathrm{ch}}
\newcommand{\ad}{\mathrm{ad}}     			

\newcommand{\tta}{\mathtt{a}}

\newcommand{\frg}{\mathfrak{g}}				

\newcommand{\frM}{\mathfrak{M}}

\begin{document}

\title[Noncommutative Instantons]{Noncommutative Instantons in Diverse Dimensions}


\author*[1,2,3]{\fnm{Richard J.} \sur{Szabo}}\email{R.J.Szabo@hw.ac.uk}

\author[1,2]{\fnm{Michelangelo} \sur{Tirelli}}\email{mt2001@hw.ac.uk}
\equalcont{This author contributed equally to this work.}

\affil*[1]{\orgdiv{Department of Mathematics}, \orgname{Heriot-Watt University}, \orgaddress{\street{Colin Maclaurin Building, Riccarton}, \city{Edinburgh} \postcode{EH14 4AS},  \country{United Kingdom}}}

\affil[2]{\orgname{Maxwell Institute for Mathematical Sciences},  \orgaddress{\street{The Bayes Centre}, \city{Edinburgh} \postcode{EH8 9BT}, \country{United Kingdom}}}

\affil[3]{\orgname{Higgs Centre for Theoretical Physics}, \orgaddress{\street{James Clerk Maxwell Building, Kings Buildings}, \city{Edinburgh} \postcode{EH9 3JZ}, \country{United Kingdom}}}


\abstract{This is a mini-review about generalized instantons of noncommutative gauge theories in dimensions~4, 6 and 8, with emphasis on their realizations in type~II string theory, their geometric interpretations, and their applications to the enumerative geometry of non-compact toric varieties. \hfill{~~}}

\keywords{Noncommutative Instantons, D-Branes, Cohomological Field Theories, Enumerative Geometry}

\pacs[Preprint]{EMPG--22--15}



\maketitle

\section{Introduction}\label{sec1}

Instantons in Yang-Mills theory on $4$-dimensional Euclidean space $\mathbbm{R}^4$ are solutions to (anti-)self-duality equations for gauge fields with suitable boundary conditions at infinity~\cite{Belavin:1975fg}. They are labelled by their topological Pontryagin charge, called the instanton number, and are absolute minima of the Yang-Mills action functional. 

In the quantum theory, instantons are of particular interest as they provide non-perturbative contributions to the path integral. However, the instanton partition functions are divergent because the instanton moduli spaces are non-compact, due to singularities where instantons shrink to zero size and also due to the non-compactness of the ambient space $\FR^4$ in which instantons can run away to infinity. 
A UV regularization which resolves the small instanton singularities is provided by deforming the equations to instanton equations in noncommutative field theory~\cite{Nekrasov:1998ss}. An IR regularization is provided by putting the gauge theory in an $\Omega$-background and regarding the instanton partition function as the corresponding equivariant integral over the instanton moduli space~\cite{SWcounting}. 

From a geometric point of view, the compactified instanton moduli space is isomorphic to a moduli space of torsion free sheaves on the complex projective plane $\mathbbm{P}^2$ with suitable characteristic classes and framing conditions~\cite{Nakajimabook}. From a physical perspective, instantons arise in type~II string theory as D$p$--D$(p{+}4)$-brane bound states in the low energy limit~\cite{Moore:1998et}. The noncommutative deformation corresponds to turning on a non-zero constant background $B$-field in the flat $10$-dimensional target spacetime~\cite{Seiberg:1999vs}. 

Yang-Mills instantons on $\FR^4$ can be extended to `generalized instantons', which are Euclidean BPS solutions of Yang-Mills theories in higher even dimensions $2n>4$ satisfying generalized (anti-)self-duality equations~\cite{Corrigan:1982th}. They are defined on spaces of special holonomy and naturally appear in type~II string theory, where one is interested in instantons in dimensions up to~$8$ ($n=4$). The corresponding instanton moduli spaces are related to various problems of interest in enumerative algebraic geometry, which has contributed to the ongoing fruitful exchange of ideas and techniques between physics and mathematics.  

It is the purpose of this mini-review to survey these and other interactions between noncommutative geometry, string theory and algebraic geometry. We present both old and recent developments from a unified perspective with emphasis on their commonalities and differences in the various dimensions $4,6,8$. Throughout we highlight the many open problems that remain, with plenty of pointers to the relevant literature for more details and further reading. In Section~\ref{sec:NCinst} we review the constructions and geometry of instantons on noncommutative $\FR^{2n}$, explaining their realisations in string theory in Section~\ref{sec:ADHM}. In Section~\ref{sec:partfns} we explain the computation of equivariant instanton partition functions and their applications to enumerative geometry. In Section~\ref{sec:further} we survey developments beyond the case of flat space $\FR^{2n}$ including noncommutative instantons on orbifolds, on defects, and on general toric Calabi-Yau manifolds.

\section{Noncommutative Instantons on $\FR^{2n}$}\label{sec:NCinst}

\subsection{Generalized Instanton Equations}\label{subsec:geninst}

Let $M_n$ be a connected oriented Riemannian manifold of even dimension $2n\geq4$, and let $\ast$ be the associated Hodge duality operator acting on differential forms on $M_n$. A (\emph{generalised}) \emph{instanton} is a localized finite-action solution to the classical Euclidean field equations of a gauge theory on $M_n$. They are solutions to first order $\Sigma$-self-duality equations~\cite{8d,8ds,Popov:2010rf}
\begin{align}
\Sigma\wedge F=\ast\, F  \label{ing}
\end{align}
for a $\sU(r)$-connection $1$-form $A$ with curvature $2$-form $F=\dd A+A\,\wedge\, A$. Here $\Sigma$ is a differential form of degree $2n-4$ which is taken to be a singlet of the holonomy group $\sH \subset \sSO(2n)$, a maximal subgroup of the structure group of the frame bundle of $M_n$ which is preserved by the equations~\eqref{ing}.

If $\Sigma$ is closed, $\dd\Sigma=0$, then the instanton solutions of \eqref{ing} are labelled by topological numbers corresponding to the Chern classes of the associated gauge bundle $\CE\rightarrow M_n$; in particular, the $n$-th Chern class ${\rm ch}_n(\CE)$ determines the instanton charge $k\in\RZ$. The Bianchi identity $\dd_AF=\dd F+A\wedge F=0$ together with \eqref{ing} imply
\begin{align}
   \dd_A \ast F = \dd_A(\Sigma\wedge F) = \Sigma\wedge \dd_A F = 0 \ .
\end{align}
In this case every solution of the instanton equations is a solution of the second order Yang-Mills equations, i.e. an extremum of the Yang-Mills action functional $S_{\textrm{\tiny YM}}$ on $M_n$. 

\noindent\framebox[1.1cm]{\small{$\boldsymbol{n=2}$}} \ 
In $4$ dimensions, $\Sigma$ is a locally constant function on $M_2$ and the maximal holonomy group is $\sH=\sSO(4)$. Since $\ast^2=\ident$ when acting on $2$-forms, in this case
\begin{align}
F= \ast\, (\Sigma \, F)=\Sigma\ast F=\Sigma^2\, F \ ,
\end{align}
and so $\Sigma=\pm\, 1$. We will work with anti-self-dual instantons~\cite{inst,Vandoren:2008xg}, satisfying $\ast\, F=- F$.

\noindent\framebox[1.1cm]{\small{$\boldsymbol{n=3}$}} \ 
In $6$ dimensions, $\Sigma$ is a closed $2$-form and the only allowed holonomy groups are $\sSO(4){\times}\sSO(2)$ and $\sU(3)$~\cite{Matrix}. The first choice corresponds to the case where $M_3$ is the direct product of a $4$-manifold and a Riemann surface. We will assume that $\Sigma$ is non-degenerate and defines a K\"ahler structure $\omega$ on $M_3$. Then $\sH=\sU(3)$.

\noindent\framebox[1.1cm]{\small{$\boldsymbol{n=4}$}} \ 
In $8$ dimensions, $\Sigma$ is a closed $4$-form and the maximal holonomy group is $\sH=\sSpin(7)\subset\sSO(8)$~\cite{Matrix,Ward:1983zm}. We will take $\Sigma$ to be a $\sSpin(7)$-structure on $M_4$, which in a local chart is given by the Cayley $4$-form~\cite{spin}. In this case the self-adjoint operator $\ast\circ(\Sigma\wedge-)$ acting on $2$-forms has eigenvalues $\lambda=-3,1$. We will take $\lambda=1$.

\subsection{Instanton Singularities}\label{subsec:Derrick}

Derrick's Theorem~\cite{Derrick:1964ww} forbids localized solutions on $M_n\!=\!\FR^{2n}$ with finite Yang-Mills action when $n>2$: If $A=A_\mu(x)\,\dd x^\mu$ is a localized solution of the Yang-Mills equations, we can deform the field $A$ by rescaling the coordinates as $x\rightarrow\lambda\, x$ for $\lambda\in\FR_{>0}$. Then the Yang-Mills functional $S_{\textrm{\tiny YM}}$ changes to $S_{\textrm{\tiny YM}}(\lambda)=\lambda^{4-2n}\, S_{\textrm{\tiny YM}}$ with \smash{$S'_{\textrm{\tiny YM}}(1) = (4-2n) \, S_{\textrm{\tiny YM}} < 0$}, and hence $A$ is no longer a localized solution.
It follows that the moduli space of solutions to the generalized instanton equations \eqref{ing} is empty. 

This problem is cured by deforming the equations \eqref{ing} to noncommutative instanton equations. Since the theory is no longer formulated on flat Euclidean space, Derrick's Theorem does not apply. The coordinates cannot be rescaled without affecting the Moyal-Weyl commutation relations $[x^\mu,x^\nu] = \ii\,\theta^{\mu\nu}$.

However, even in $4$ dimensions, we have to contend with the small instanton UV singularities. For example, let $\eta^a_{\mu\nu}$ for $a=1,2,3$ be the ’t~Hooft symbols and $\sigma_a$ the generators of the Lie algebra $\mathfrak{su}(2)$. An $\sSU(2)$ instanton solution $A=A^a\otimes\sigma_a$ on $\mathbb{R}^4$ is given by the celebrated family of BPST instantons \cite{tHooft:1976snw}
\begin{align}
    A^a=2 \, \frac{\eta_{\mu\nu}^a \,(x-x_0)^\nu\,\dd x^\mu}{\vert  x-x_0\vert ^2+\lambda^2} \ ,
\end{align}
which is parametrized by the center of mass modulus $x_0\in\FR^4$ and the instanton scale $\lambda\in\FR_{>0}$. This solution becomes singular when its scale is shrunk to zero size ($\lambda\rightarrow0$). 

These singularities are again resolved by considering noncommutative instantons. This introduces a minimum size $\xi$ for the instantons, and so it resolves the moduli space singularities as instantons cannot reach the singularity where their size vanishes. Because of the additional length scale $\xi$, there is no commutative equivalent of this resolution of singularities.

Finally, for any $n\geq2$, we still have to contend with the non-compact IR singularities which arise from instantons that run away to infinity in $\FR^{2n}$. These are resolved by coupling the gauge theory to Nekrasov’s supergravity $\Omega$-background~\cite{SWcounting}.

\subsection{Hermitian Yang-Mills Equations}
\label{subsec:HYM}

The noncommutative deformation is equivalent to the choice of a complex structure $J$ on $M_n=\mathbbm{R}^{2n}$, which induces a Poisson bivector $\theta=\xi\,\omega^{-1}$ that we wish to quantize, where $\xi\in\FR_{>0}$ and $\omega$ is the corresponding K\"ahler $(1,1)$-form of $M_n$. The instanton equations are now generally defined on the K\"ahler $n$-fold~$(M_n,\omega,J)$, and under suitable conditions they are equivalent to the Hermitian Yang-Mills equations
\begin{align}\label{eq:HYM}
F^{0,2} = 0   \quad ,   \quad \omega^{n-1}\wedge F^{1,1} = \mu \ \omega^n \,\ident_r \ ,
\end{align}
where $F=F^{2,0}+F^{1,1}+F^{0,2}$ is the decomposition of the field strength in the standard basis of $(1,0)$- and $(0,1)$-forms, and $\mu$ is a constant which is non-zero only when the gauge bundle $\CE$ has non-trivial first Chern class. The first equation turns $\CE$ into a holomorphic bundle, while the second equation can be regarded as a stability condition. 

We will usually consider gauge fields with vanishing first Chern form $\ch_1(\CE)=0$, or alternatively we may consider only the traceless part of the field strength tensor $F$, and hence assume $\mu=0$. In this case the equations \eqref{eq:HYM} coincide with the Donaldson-Uhlenbeck-Yau (DUY) equations~\cite{Donaldson:1985zz,UhlenbeckYau} which describe stable holomorphic vector bundles over $M_n$ with finite characteristic classes.

\noindent\framebox[1.1cm]{\small{$\boldsymbol{n=2}$}} \ 
In $4$ dimensions, the introduction of a complex structure $J$ breaks the holonomy group to $\sU(2)\subset\sSO(4)$, and the anti-self-duality equations $\ast\,F=- F$ are equivalent to the Hermitian Yang-Mills equations \eqref{eq:HYM} for $n=2$ and $\mu=0$.

\noindent\framebox[1.1cm]{\small{$\boldsymbol{n=3}$}} \ 
In $6$ dimensions, the holonomy group is similarly $\sU(3)\subset\sSO(6)$, and the generalized instanton equations $\omega\wedge F=\ast\, F$ are again equivalent to \eqref{eq:HYM} for $n=3$ and $\mu=0$.

\noindent\framebox[1.1cm]{\small{$\boldsymbol{n=4}$}} \ 
In $8$ dimensions, $J$ defines a Calabi-Yau structure and reduces the holonomy to $\sSU(4)\subset\sSpin(7)$. A compatible $\sSpin(7)$-structure is determined by $\Sigma=\frac12\,\omega\wedge\omega - {\rm Re}(\Omega)$, where $\Omega$ is the non-degenerate holomorphic $4$-form associated to the $\sSU(4)$-structure. The complex conjugate of $\Omega$ defines an antilinear involution $\ast_\Omega$ acting on $(0,2)$-forms on $M_4$, which enables one to introduce the anti-self-dual part $
F_-^{0,2}=\tfrac12\,\big(F^{0,2}-\ast_\Omega\, F^{0,2}\big)
$
of $F^{0,2}$ in the $-1$ eigenspace of $\ast_\Omega$~\cite{Donaldson:1996kp,Donaldson:2009yq}. The $\sSpin(7)$-instanton equation $\ast\,(\Sigma\wedge F) = F$ can be reduced along the inclusion $\sSU(4)\subset\sSpin(7)$ to 
\begin{align}\label{Insteq_d8}
    F_{-}^{0,2}=0   \quad ,   \quad \omega\wedge\omega\wedge\omega\wedge F^{1,1} =0 \ .
\end{align}
Any solution of the Hermitian Yang-Mills equations \eqref{eq:HYM} for $n=4$ and $\mu=0$ is automatically a solution of the $\sSpin(7)$-instanton equations \eqref{Insteq_d8}, but not conversely unless the topological number $\int_{M_4} \,\Omega\wedge \Tr\big(F^{0,2}\wedge F^{0,2} \big)$ vanishes~\cite{inprep}.

\subsection{Noncommutative Gauge Theory}\label{subsec:NCinsteq}

The noncommutative deformation of the generalized instanton equations on the K\"ahler manifold $M_n=\FR^{2n}\simeq\FC^n$ uses Berezin-Toeplitz quantization with the Poisson structure $\theta = \xi\,\omega^{-1}$. Starting from the trivial prequantum line bundle $L\rightarrow M_n$ with holomorphic polarization, the Hilbert space $\CH = H^0(M_n,L)=\ker\big(\bar\partial_\alpha\big)$ is the space of holomorphic sections of $L$, where $\omega=\xi\,\dd\alpha$. Holomorphic functions on $M_n$ are naturally realized as operators on $\CH$. 

In particular, the Toeplitz quantization map sends the local complex coordinates $\{z_a,\bar{z}_{\bar{a}}\}_{a=1}^n$ of $\FC^n$ to operators with the commutation relations
\begin{align}\label{com_relations}
    [z_a,z_b]=0=[\bar{z}_{\bar{a}},\bar{z}_{\bar{b}}] \  , \ [z_a,\bar{z}_{\bar{b}}]=\xi\,\delta_{a\bar{b}} \, \ident_\CH \ .
\end{align}
By the Stone-von~Neumann theorem, the Hilbert space $\CH$ is isomorphic to the unique irreducible representation of this algebra, the Fock module
\begin{align}\label{eq:Fock}
\begin{split}
\CH = \CH_{1\cdots n} &= \FC[\bar z_1,\dots,\bar z_n]\,\vert {\vec 0}\,\rangle \\[4pt]
&= \mbox{$\bigoplus\limits_{{\vec n}\in\RZ_{\geq0}^n}$} \, \FC\,\vert {\vec n}\rangle \ ,
\end{split}
\end{align}
where the vacuum vector $\vert \vec 0\,\rangle$ is a fixed section in $\ker(\bar\partial_\alpha)$ and $\vec n=(n_1,\dots,n_n)$. The operators $\bar z_{\bar a}$ and $z_a$ act on $\CH$ as creation and annihilation operators, respectively,
and so $\CH$ is spanned by the eigenstates $\vert \vec n\,\rangle$ of the number operators $N_a=\frac1{\xi} \, \bar z_{\bar a}\,z_a$ with eigenvalues $n_a$ for $a=1,\dots,n$. 

Using the Segal-Bargmann representation \smash{$\bar z_{\bar a}=-\xi\,\frac\partial{\partial z_a}$} of the algebra \eqref{com_relations}, the $\sU(r)$ instanton equations \eqref{eq:HYM} (with $\mu=0$) and \eqref{Insteq_d8} can now be rewritten as equations for infinite-dimensional matrices $Z_a=\ident_r\otimes \bar z_{\bar a} + \ii\,\xi\,(A_{2a-1}+\ii\,A_{2a})$ acting on $\CH^r = \FC^r\otimes\CH$. They are algebraic operator equations for the noncommutative fields given by
\begin{align}\label{alg_insteqd8}
\begin{split}
[Z_a,Z_b] &= \left\{ \begin{matrix}
0 \ , & n=2,3 \\[0.5em] \tfrac{1}{2}\,\epsilon_{ab\bar c\bar d}\,\big[Z^\dagger_{\Bar{c}},Z^\dagger_{\Bar{d}}\big] \ , & n=4 
\end{matrix} \right.  \ , \\[4pt]
\mbox{$\sum\limits_{a=1}^n$} & \, \big[Z_{\Bar{a}}^\dagger,Z_a\big] = -n \, \xi \, \ident_{\CH^r} \ .
\end{split}
\end{align}
These equations are much easier to analyse and solve than the original first order partial differential equations.

\subsection{Explicit Solutions}\label{subsec:solutions}

The simplest solution to the equations \eqref{alg_insteqd8} is the vacuum solution \begin{align}\label{eq:vacsol}
Z_a=\ident_r\otimes\bar z_{\bar a}
\end{align}
with vanishing gauge connection. More general solutions can be found using the solution generating technique which applies partial gauge transformations to the vacuum solution \eqref{eq:vacsol}~\cite{Harvey:2000jb,Hashimoto:2000kq,Gross:2000ss,Kraus:2001xt,Nekrasov:2002kc}.  Explicit finite action BPS and non-BPS solutions of the noncommutative DUY equations on $\FR^{2n}$ for any $n\geq1$ were studied extensively in~\cite{Lechtenfeld:2003cq,Popov:2005ik,Lechtenfeld:2006wu,Lechtenfeld:2007st,Lechtenfeld:2008nh}. 

\noindent{\small\underline{\bf NS-Type Instantons.}} \ 
The Nekrasov-Schwarz \smash{(NS)} type instantons include the instanton solutions that were originally found in~\cite{Nekrasov:1998ss,Furuuchi:1999kv} via a noncommutative version of the ADHM construction. Focusing momentarily on the rank~$1$ case $r=1$, we fix an integer $\ell>0$ and a partial isometry $S_\ell$ which projects out the finite-dimensional subspace of states $\vert  \vec n\, \rangle$ with $\vert \vec n\vert :=\sum_{a=1}^n\,n_a<\ell$ from the Fock space $\CH$. Explicitly
\begin{align}\label{eq:Uell}
    S_\ell^\dagger\, S_\ell=\ident_\CH-\varPi_\ell   \quad ,   \quad S_\ell\,S_\ell^\dagger=\ident_\CH \ ,
\end{align}
with $\varPi_\ell=\sum_{\vert \vec n\vert <\ell}\,\vert  \vec n\rangle\langle \vec n \vert $.
Then the NS-type instantons are given by
\begin{align}\label{eq:NStype}
    Z_a=S_\ell\,\bar z_{\bar a}\, f_\ell(N)\, S_\ell^\dagger \ ,
\end{align}
where $N=\sum_{a=1}^n\,N_a$ is the total number operator,
and $f_\ell$ is a real function of $N$ given by
\begin{align}
\begin{split}
    f_\ell(N)&=\sqrt{1-\tfrac{\ell\,(\ell+1)\cdots (\ell+n-1)}{(N+1)\cdots (N+n)}} \ (\ident_\CH-\varPi_\ell) \ .
    \end{split}
\end{align}

These solutions have finite Yang-Mills action and topological charge
\begin{align}
\begin{split}
k = \ch_n(\CE) &= \frac{(-\pi\,\xi)^n}{n!} \, \Tr_\CH\Big(\frac F{2\pi}\Big)^n \\[4pt]
&=\Tr_\CH(\varPi_\ell) = \binom{\ell+n-1}{\ell-1} \ ,
\end{split}
\end{align}
which is the number of states removed by the shift operator $S_\ell$. They correspond to localized instantons sitting outside a characteristic radius $\sqrt{\ell\,\xi}$ from the origin in $\FR^{2n}$. This demonstrates how the noncommutative deformation resolves the singular configuration space of commutative instantons which would sit on top of each other. Higher rank analogues of these solutions are described in~\cite{Popov:2005ik,Ivanova:2006ek,Cirafici:2008sn}.

\noindent{\small\underline{\bf ABS Construction.}} \ 
By relaxing the conditions \eqref{eq:Uell} one can generate more general solutions of higher rank $r>1$ with instanton charge $k$ by appealing to the Toeplitz algebra of the space $\CH^r=\FC^r\otimes\CH$. A Toeplitz operator is a partial isometry $T_k:\CH^r\rightarrow\CH^r$ with trivial kernel and $k$-dimensional cokernel, that is, it obeys
\begin{align}
T_k^\dag\,T_k = \ident_{\CH^r}   \quad ,   \quad T_k\,T_k^\dag = \ident_{\CH^r} - P_k \ ,
\end{align}
where $P_k$ is a rank $k$ projector on the space $\CH^r$. Then the basic shift operator soliton is given by $
Z_a = T_k\,(\ident_r\otimes\bar z_{\bar a}) \, T_k^\dag $.

An explicit realization of these Toeplitz operators for rank $r=2^{n-1}$ is given by the noncommutative ABS construction~\cite{Harvey:2000te}. We use real polarization and set $T_k=(T)^k$ with
\begin{align}\label{eq:ABS}
T^\dag = \tfrac1{\sqrt{(\gamma\cdot x)\,(\gamma\cdot x)^\dag}} \, \gamma\cdot x \ ,
\end{align}
where $\gamma\cdot x=\gamma_\mu\,x^\mu$ and the $r{\times}r$ matrices $\gamma_\mu$ obey anticommutation relations such that 
\begin{align}
\Gamma_\mu={\small\bigg( \begin{matrix} 0 & -\gamma_\mu \\ \gamma_\mu^\dag & 0\end{matrix}} \normalsize\bigg)
\end{align} 
generate the Clifford algebra of the inner product space $(\FR^{2n},\delta_{\mu\nu})$. The operator $T$ has trivial kernel, while $T^\dag$ has a $1$-dimensional kernel which is spanned by the vector $\vert \alpha\rangle\otimes\vert \vec0\,\rangle$, where $\vert \alpha\rangle$ denotes the lowest weight spinor of $\sSO(2n)$. The ABS construction \eqref{eq:ABS} can be straightforwardly generalized in order to introduce $2\,n\,k$ real moduli into the solution which specify the locations of the $k$ noncommutative instantons on $\FR^{2n}$~\cite{Lechtenfeld:2003cq}.

\subsection{Moduli Spaces of Instantons}\label{subsec:Hilbertscheme}

The explicit solutions obtained in Section~\ref{subsec:solutions} elucidate the geometric description of the instanton moduli spaces $\frM_{k,r}(\FC^n)$ for $n=2,3,4$, that is, the (suitably compactified) quotient space of charge~$k$ solutions to the rank~$r$ instanton equations \eqref{eq:HYM} (with $\mu=0$) and \eqref{Insteq_d8} on $\FR^{2n}$ by gauge transformations. Focusing momentarily on the rank~$1$ case, there is a correspondece between $\sU(1)$ instantons of charge $k$ on noncommutative $\FR^{2n}$ and ideals $\CI$ of codimension $k$ in the polynomial ring $\FC[w_1,\dots, w_n]$: In all of the solutions discussed in Section~\ref{subsec:solutions}, the partial isometries bijectively identify the Fock space $\CH$ with a subspace
\begin{align}\label{eq:CHCI}
    \CH_\CI=\CI[\bar z_1,\dots,\bar z_n]\vert \vec0\,\rangle \ ,
\end{align}
where $\CI$ is the ideal of codimension $k$ formed by the polynomials $f\in\FC[w_1,\dots,w_n]$ satisfying
\begin{align}\label{eq:Pkf}
    P_k f(\bar z_1,\dots,\bar z_n)\vert \vec{0}\,\rangle=0 \ .
\end{align}
These ideals parametrize the Hilbert scheme $\Hilb^k(\FC^n)$ of $k$ points in $\FC^n$, and hence there is an isomorphism
\begin{align}
\frM_{k,1}(\FC^n) \, \simeq \, \Hilb^k(\FC^n) \ .
\end{align}

Geometrically, $\Hilb^k(\FC^n)$ parametrizes $0$-dimensional subschemes $Z\subset\FC^n$ with Hilbert-Poincar\'e polynomial $h^0(\CO_Z) = k$. For $n=2$ and generic $k$, the Hilbert scheme $\Hilb^k(\FC^2)$ is a smooth manifold of complicated topology~\cite{Nakajimabook}.
For $n=3,4$ and $k>3$, however, the Hilbert scheme is not smooth: it is not even a manifold, as in general it has several different branches of varying dimension. 

In the general higher rank case $r>1$, we note that the vacuum solution \eqref{eq:vacsol} in noncommutative gauge theory is a realization of the free coherent sheaf of sections of the trivial rank $r$ holomorphic vector bundle $\CO^{\oplus r}$ on $\FC^n$. Then each choice of partial isometry with the identifications \eqref{eq:CHCI} and \eqref{eq:Pkf} realises a $0$-dimensional quotient $\CO^{\oplus r} \twoheadrightarrow\CO_Z$ of $\CO^{\oplus r}$ with $h^0(\CO_Z) = k$. These parametrize the Quot scheme $  \Quot^k_{\FC^n}(\CO^{\oplus r})$ of $k$ points in $\FC^n$, and hence there is an isomorphism
\begin{align}
\frM_{k,r}(\FC^n) \, \simeq \,   \Quot^k_{\FC^n}(\CO^{\oplus r}) \ .
\end{align}
For $r=1$ this is the same as the Hilbert scheme: $  \Quot^k_{\FC^n}(\CO) = \Hilb^k(\FC^n)$. 

The same moduli space $  \Quot^k_{\FC^n}(\CO^{\oplus r})$ also parametrizes framed torsion free sheaves $\CE$ on projective space $\PP^n$ of rank $r$ and $\ch_n(\CE)=k$~\cite{Cazzaniga:2020xru}: Any such sheaf sits in a short exact sequence
\begin{align}
0\longrightarrow \CE\longrightarrow \CO_{\PP^n}^{\oplus r}\longrightarrow \CS_Z \longrightarrow 0 \ ,
\end{align}
where $\CS_Z=\iota_*\CO_Z$ is a coherent sheaf supported on a set of $k$ points $Z\subset\FC^n$ under the canonical open embedding $\iota:\FC^n \hookrightarrow \PP^n$. For $n=2$, these are the celebrated Gieseker-Nakajima moduli spaces $\frM_{k,r}(\FC^2)$ which parametrize charge $k$ noncommutative $\sU(r)$ instantons on $\FR^4$~\cite{Nekrasov:1998ss}, framed rank~$r$ torsion free sheaves $\CE$ on $\PP^2$ with ${\rm ch}_2(\CE)=k$~\cite{Nakajimabook}, and framed algebraic bundles on a noncommutative projective plane~\cite{Kapustin:2000ek}.

\section{Noncommutative Instantons from String Theory}\label{sec:ADHM}

\subsection{Cohomological Gauge Theory}\label{subsec:CohFT}

The Hermitian Yang-Mills equations have a long history in string theory dating back to the mid-1980s, where they naturally appear as the conditions for survival of at least one unbroken supersymmetry in the low-energy effective field theory in $d\leq4$ dimensions. They have acquired renewed interest due to their relevance in the description of D-branes in type~IIA string theory. The low-energy effective dynamics on a stack of $r$ D-branes can be obtained by dimensional reduction from $10$-dimensional $\mathcal{N}=1$ supersymmetric Yang-Mills theory with gauge group $\sU(r)$, which after topological twisting become cohomological field theories~\cite{Blau:1997pp}. Generalized instanton equations then appear as the BRST fixed point loci on which these theories localize.

We start by compactifying type~IIA string theory on
\begin{align}\label{compactification}
\FR^{1,9-2n}\times M_n  \  , 
\end{align}
where $M_n$ is a K\"ahler $n$-fold, and consider the (twisted) low-energy effective field theories on $r$ D$(2n)$-branes wrapping the space $M_n$~\cite{Bershadsky:1995qy}. The gauge theory is defined on $M_n$, while holomorphic coordinates of $\FR^{1,9-2n}$ become complex scalar fields of the theory. The original Lorentz symmetry $\sSO(1,9)$ in ten dimensions is broken generically to $\sSO(1,9-2n)\times\sU(n)$. In the bosonic sector, the non-compact R-symmetry $\sSO(1,9-2n)$ acts only on the scalars. 

Instantons of charge $k=\ch_n(\CE)$ in this theory are realised as supersymmetric bound states of $k$ D0-branes inside the D$(2n)$-branes~\cite{Szabo:2004uy,Polchinski:1998rr}. More generally, the Chern forms $\ch_p(\CE)$ couple to Ramond-Ramond $(2n{-}2p)$-form fields on $M_n$ with $1\leq p\leq n$ and represent Pontryagin charges of D$(2n{-}2p)$--D$(2n)$ bound states.

\noindent\fbox{\small{$\boldsymbol{n=2}$}} \ 
We reduce $10$-dimensional supersymmetric Yang-Mills theory on a $4$-dimensional K\"ahler manifold $(M_2, \omega)$  with $\sU(2)$ holonomy. After the Vafa-Witten twist~\cite{Vafa:1994tf}, the bosonic field content consists of a gauge field $A$, a self-dual $2$-form $B^+$ which is an auxiliary field, and a complex Higgs field $\Phi$, all valued in the adjoint representation of the gauge group $\sU(r)$. The bosonic action is minimized by the equations 
\begin{align}\label{eq_red_d4}
\begin{split}
F^++\tfrac{1}{8}\,[B^+ \,{\scriptstyle{\mathbf\times}}\,
B^+]+\tfrac{1}{2}\,[B^+,\Phi]&=0 \ , \\[4pt] \dd^\ast_AB^++\dd_A \Phi&=0 \ , 
\end{split}
\end{align}
where $F^+=\frac12\,(F+\ast\,F)$ is the self-dual part of the curvature $2$-form. We will only consider solutions of the partial differential equations \eqref{eq_red_d4} which have $B^+=0$, yielding $4$-dimensional instantons, together with parallel Higgs fields that will play an important role below.

\noindent\fbox{\small{$\boldsymbol{n=3}$}} \ 
We similarly reduce the theory in ten dimensions on a $6$-dimensional K\"ahler manifold $(M_3, \omega)$  with $\sU(3)$ holonomy. After twisting, the bosonic spectrum consists of a gauge field $A$, a $(3,0)$-form $\varrho$ and a complex Higgs field $\Phi$, all valued in the adjoint representation of $\sU(r)$. The twisted gauge theory corresponds to the moduli problem associated with the field equations~\cite{Baulieu:1997jx,Iqbal:2003ds}
\begin{align}\label{eq_red_d6}
\begin{split}
F^{2,0} + \bar{\partial}_A^\ast\,\varrho &= 0 \ , \\[4pt]
\omega\wedge\omega\wedge F^{1,1} + \tfrac12\,[\varrho,\bar{\varrho}] &= \mu \ \omega\wedge\omega\wedge\omega \ \ident_r \ , \\[4pt]
\dd_A\Phi&=0 \ . 
\end{split}
\end{align}
We will always assume $\mu=0$, i.e. $\ch_1(\CE)=0$, which in the IIA string picture excludes D$4$--D$6$ bound states. If we further restrict to $\sSU(3)$ holonomy, the Calabi-Yau structure enables us to set $\varrho=0$ because of uniqueness of the holomorphic $3$-form in that case. Then the first two equations of \eqref{eq_red_d6} reduce to the DUY equations.

\noindent\fbox{\small{$\boldsymbol{n=4}$}} \ 
Dimensional reduction of $10$-dimensional $\CN=1$ supersymmetric Yang-Mills theory on a manifold $M_4$ of $\sSpin(7)$-holonomy is equivalent to a topological twist of the resulting $8$-dimensional supersymmetric gauge theory~\cite{Baulieu:1997jx}. The bosonic field content of the reduced theory on the K\"ahler $4$-fold $(M_4,\omega)$, with holonomy $\sSU(4)\subset\sSpin(7)$, consists of a complex gauge field $A$ and adjoint Higgs field $\Phi$, and the path integral localizes onto the moduli space of solutions of the equations~\cite{inprep}
\begin{align}\label{eq_red_d8}
\begin{split}
F^{0,2}_-&=0 \ , \\[4pt]
\omega\wedge\omega\wedge\omega\wedge F^{1,1} &= \mu \ \omega\wedge\omega\wedge\omega\wedge\omega \ \ident_r \ , \\[4pt] 
\dd_A\Phi&=0 \ .
\end{split}
\end{align}
Setting $\mu=0$ amounts to excluding D$6$--D$8$ bound states, in which case the first two equations coincide with the $\sSpin(7)$-instanton equations \eqref{Insteq_d8}. We may further exclude D$4$--D$8$ bound states by restricting to solutions which have vanishing charge $\int_{M_4} \,\Omega\wedge \Tr\big(F^{0,2}\wedge F^{0,2} \big)=0$, in which case \eqref{eq_red_d8} also reduce to the DUY equations.

Noncommutative gauge theories arise from the dynamics of open strings on D-branes in the presence of a large Kalb-Ramond field on $M_n=\FR^{2n}$~\cite{Seiberg:1999vs}, which allows for a unified treatment of worldvolume theories of D-branes of various dimensions~\cite{Douglas:2001ba,Szabo:2001kg}. In addition to the algebraic instanton equations \eqref{alg_insteqd8}, the Higgs fields $\Phi$ obey
\begin{align}\label{eq:ZPhi}
[Z_a,\Phi] = 0 = [Z_{\bar a}^\dag,\Phi] \ ,
\end{align}
for $a=1,\dots,n$.

\subsection{ADHM Quiver Matrix Model}\label{subsec:Dbranes}

\begin{figure}[h]%
\centering
\includegraphics[width=0.3\textwidth]{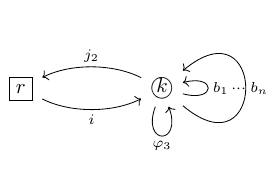}
\caption{The ADHM quiver. The arrow $j_2$ (resp. $\varphi_3$) is non-zero only for $n=2$ (resp. $n=3$). Vertices labelled by $r$ and $k$ are associated to $r$ D$(2n)$-branes and $k$ D$0$-branes, respectively. The arrows $i,j_2$ represent the D$0$--D$(2n)$ strings which carry fields in the bifundametal representation of $\sU(r)\times \sU(k)$, while $\varphi_3,b_1,\dots,b_n$ describe D$0$--D$0$ strings which generate (complex) $\sU(k)$ gauge fields.
}\label{fig:ADHMquiver}
\end{figure}

The equations \eqref{alg_insteqd8} for $\sU(r)$ noncommutative instantons of charge $k$ describe the low-energy interactions of $k$ D$0$-branes with $r$ D$(2n)$-branes in type~IIA string theory in a large Kalb-Ramond field, from the perspective of the worldvolume gauge theory on the D$(2n)$-branes. On the other hand, from the perspective of the D$0$-branes, wherein the D$(2n)$-branes are heavy, the bound states are described by (generalised) ADHM equations, deformed by a Fayet-Iliopoulos coupling $\xi$ related to the non-zero $B$-field~\cite{Nekrasov:1998ss,Ohta:2001dh,Hiraoka:2002wm,Cirafici:2008sn,Nekrasov:2017cih,Nekrasov:2018xsb,Bonelli:2020gku,inprep}: introducing noncommutativity corresponds to adding a Fayet-Iliopoulos term $\xi$ in string theory. This gives a physical derivation of the ADHM construction of instantons~\cite{inst}: the low-energy effective theory on the D$0$-branes is a cohomological matrix model~\cite{Moore:1997dj,Moore:1998et}, whose field content can be succinctly encoded in a representation of a quiver, called the (generalized) ADHM quiver (see Figure~\ref{fig:ADHMquiver}). This perspective in the various dimensionalities is nicely reviewed in~\cite{Kanno:2020ybd}.

The BRST fixed points of the cohomological matrix model yield relations for the ADHM quiver, together with D-term conditions. These are called ADHM equations and they can be identified with the vacuum equations of the supersymmetric gauge theory. This gives an algebraic description of the instanton moduli space $\frM_{k,r}(\FC^n)$ as a moduli space of stable representations of the bounded ADHM quiver, which is the moduli space of vacua of the Higgs branch from the perspective of the D$0$-brane theory.

The case $n=2$ is special. In $4$ dimensions the ADHM quiver is the double of the framed Jordan quiver, and the choice $\xi=0$ is permitted, which corresponds to turning off the $B$-field. This is why commutative instantons exist on $\FR^4$. However, in this case the small instanton singularity persists: it corresponds to the transition between the Coulomb and Higgs branches of the D$0$-brane theory~\cite{Eto:2004rz,Closset:2020scj}, where the D$0$-branes escape from the D$4$-branes. 

In marked contrast, choosing $\xi=0$ when $n=3,4$ would reduce the open string spectrum to only the D$0$--D$0$ strings, which does not describe instantons. In other words, instanton solutions on $\FR^6$ or $\FR^8$ only exist for $\xi>0$: there are no stable BPS bound states of D$0$--D$(2n)$-branes in $2n$ dimensions for $n=3,4$ unless a (supersymmetry preserving) Kalb-Ramond field is turned on~\cite{Witten:2000mf}.

\subsection{Instanton Partition Functions}

The local geometry of the moduli space of $\sU(r)$ instantons on the K\"ahler $n$-fold $M_n$ is captured by the $3$-term instanton deformation complex, which for $n=2,3,4$ respectively reads as
\begin{align}
& \Omega^0(M_2,\frg) \xrightarrow{\bar\partial_A} \Omega^{0,1}(M_2,\frg) \xrightarrow{\bar\partial_A} \Omega^{0,2}(M_2,\frg) \nonumber \\[4pt]
& \Omega^{0}(M_3,\frg) \xrightarrow{\!\binom{\bar\partial_A}{0}\!} \begin{matrix} \Omega^{0,1}(M_3,\frg)\\ \oplus \\ \Omega^{0,3}(M_3,\frg)\end{matrix}\xrightarrow{ (\bar\partial_A \ \bar\partial_A^{\ast})} \Omega^{0,2}(M_3,\frg) \nonumber \\[4pt]
& \Omega^{0}(M_4,\frg) \xrightarrow{ \bar\partial_A} \Omega^{0,1}(M_4,\frg) \xrightarrow{ \bar\partial_A^-}\Omega_-^{0,2}(M_4,\frg) \label{eq:defcomplex:d4}
\end{align}
where $\frg:=\mathfrak{gl}(r,\FC)$. The first arrows are infinitesimal complex gauge transformations, while the second arrows are the linearizations of the first of the equations in \eqref{eq_red_d4}--\eqref{eq_red_d8}. For $n=2$ we set $B^+=0$ in \eqref{eq_red_d4}, while for $n=4$ we defined $\bar\partial_A^-:=P_\Omega^-\circ\bar\partial_A$
where  $P_\Omega^-=\frac12\,(\ident-\ast_\Omega)$ is the projection to the $-1$ eigenspace \smash{$\Omega^{0,2}_-(M_4,\frg)$} of the involution $\ast_\Omega$ from Section~\ref{subsec:HYM}.

The supersymmetric field theory provides an integral representation of the enumerative moduli problem captured by these elliptic complexes, whose cohomology groups are spanned by the fermionic zero modes of the supermultiplets of the cohomological gauge theory. The contribution to the path integral of a given point in the moduli space is non-zero if the Yukawa interactions saturate these zero modes. The degree~$0$ cohomology describes infinitesimal automorphisms, which we assume vanishes by restricting to irreducible connections. The degree~$1$ cohomology describes infinitesimal deformations and represents the tangent space to the moduli space at a fixed holomorphic connection $A$. The degree~$2$ cohomology parametrizes obstructions to deformations, which is spanned by the antighost zero modes and defines the obstruction bundle over the moduli space. It represents a virtual fundamental class, and its presence means that only the virtual tangent bundle, and not the stable tangent bundle on the coarse moduli space, is well-defined. One may (very roughly) think of the virtual fundamental class as the Poincar\'e dual of the Euler class of the obstruction bundle; its degree is the difference between the ranks of the tangent and obstruction bundles, called the virtual dimension of the moduli space. The BRST symmetry localizes the path integral (in a fixed topological sector) to an integral over this cycle.

When $M_n=\FC^n$ we can stratify the moduli space into its connected components $\frM_{k,r}(\FC^n)$ labelled by the instanton number $k\in\RZ_{\geq0}$. The Atiyah-Singer index theorem computes their virtual dimensions as Euler characters of the deformation complexes \eqref{eq:defcomplex:d4} with the results
\begin{align}
\begin{split}
\textrm{vdim} \, \frM_{k,r}(\FC^n) =\begin{cases}
2\,r\,k \ ,  \quad &n=2 \\
0 \ ,  \quad &n=3 \\
r\,k \ ,  \quad &n=4 \end{cases} \ .
\end{split}
\end{align}
Partition functions of the topological field theory on $\FC^n$ are then given by integrating the Euler class of a `matter bundle' $\CCK_{k,r}\rightarrow\frM_{k,r}(\FC^n)$ whose rank equals the virtual dimension of the moduli space. The instanton paritition function is formally the generating function for the Euler characteristics of $\CCK_{k,r}$ and has the form
\begin{align}\label{eq:pf}
Z^r_{\FC^n}(q) = \sum_{k=0}^\infty \, q^k \ \int_{[\frM_{k,r}(\FC^n)]^{\rm vir}} \  e(\CCK_{k,r}) \ ,
\end{align}
where the counting parameter $q$ weighs the instanton number and is determined by the UV gauge coupling.

\noindent\fbox{\small{$\boldsymbol{n=2}$}} \ 
The first complex in \eqref{eq:defcomplex:d4} has trivial obstruction bundle, and so the stable tangent bundle $T\frM_{k,r}(\FC^2)$ is well-defined. The instanton moduli space $\frM_{k,r}(\FC^2)$ can be constructed as a smooth variety~\cite{Nakajimabook,Nakajima:2003pg}, and a natural choice for the matter bundle is the tangent bundle itself: $\CCK_{k,r}=T\frM_{k,r}(\FC^2)$. This corresponds to the coupling of topologically twisted $\CN=2$ gauge theory to an adjoint hypermultiplet, and \eqref{eq:pf} defines the Vafa-Witten partition function of $\CN=4$ supersymmetric Yang-Mills theory in $4$ dimensions.

Another canonical choice is the matter bundle $\CCK_{k,r}=\CCV_{k,r}\otimes\FC^{2r}$, where $\CCV_{k,r}$ is the `natural bundle' on the instanton moduli space $\frM_{k,r}(\FC^2)$. Its fibre over a gauge orbit $[A]$ is the vector space $\FC^k$ that features in the ADHM construction of Section~\ref{subsec:Dbranes}, on which the D$0$--D$0$ strings act. In this case \eqref{eq:pf} is the instanton partition function of $\CN=2$ supersymmetric Yang-Mills theory on $\FR^4$ coupled to $2r$ flavours of fundamental fermions.

\noindent\fbox{\small{$\boldsymbol{n=3}$}} \ 
The second complex in \eqref{eq:defcomplex:d4} defines a perfect obstruction theory~\cite{Behrend:1996fey,Li}, which has been used to construct a virtual fundamental class $[\frM_{k,r}(\FC^3)]^{\rm vir}$ in~\cite{Donaldson:1996kp,Thomas:1998uj}. In this case the partition function \eqref{eq:pf} is only defined for the pure $\CN=2$ cohomological gauge theory in $6$ dimensions~\cite{Hofman:2000yx,Iqbal:2003ds}: we simply integrate $1$ over $[\frM_{k,r}(\FC^3)]^{\rm vir}$ to get the generating function for the (virtual) `volumes' of the instanton moduli spaces. For further discussion see~\cite{Cirafici:2012qc}.

\noindent\fbox{\small{$\boldsymbol{n=4}$}} \ 
The self-dual obstruction bundle restricts to real vector bundles \smash{$\Ob_{k,r}^-\rightarrow\frM_{k,r}(\FC^4)$}. The obstruction theory defined by the third complex in \eqref{eq:defcomplex:d4} has been used to construct a virtual fundamental class $[\frM_{k,r}(\FC^4)]_{\mathfrak o}^{\rm vir}$ using locally free sheaves and perfect obstruction theory in~\cite{Cao:2014bca,Cao:2015gra}; more general constructions use derived geometry~\cite{Borisov:2015vha} or algebraic cycles~\cite{Oh:2020rnj}. It
depends on a choice $\mathfrak o$ of orientation of $\Ob_{k,r}^-$~\cite{Nekrasov:2017cih}. The stable tangent bundle is not well-defined, but a natural choice of matter bundle on $\frM_{k,r}(\FC^4)$ is again obtained from the ADHM description as $\CCK_{k,r}=\CCV_{k,r}\otimes\FC^r$; it represents the coupling of the pure $\CN=2$ cohomological gauge theory in $8$ dimensions to $r$ flavours of fundamental fermions, which in the string theory picture of Section~\ref{subsec:Dbranes} is equivalent to adding $r$ anti-D$8$-branes. The addition of anti-branes modifies neither the ADHM description nor the Fayet-Iliopoulos coupling $\xi$~\cite{inprep}.

\section{Equivariant Gauge Theory}\label{sec:partfns}

\subsection{$\Omega$-Deformation}

The integrals \eqref{eq:pf} as they stand are not generally well-defined because the non-compactness of the ambient variety $\FC^n$ generally precludes the mentioned constructions of virtual fundamental cycles to integrate over. A particularly powerful approach uses equivariant localization to define them via equivariant integrals \smash{$\oint_{\,[\frM_{k,r}(\FC^n)]^{\rm vir}}$} over the instanton moduli spaces.

The global symmetry group of the cohomological field theory is 
\begin{align}
\sG_n = \begin{cases}
\sU(r)_{\rm col}\times\sU(n) \ ,   \ n=2,3\\
\sU(r)_{\rm col} \times  \sSU(4) \times\sU(r)_{\rm fla} \ ,  \ n=4
\end{cases}
\end{align}
where the colour symmetry $\sU(r)_{\rm col}$ acts by rotating the framing of the gauge bundle at infinity in $\FR^{2n}$, and the flavour symmetry $\sU(r)_{\rm fla}$ acts on its vector representation $\FC^r$. (When the $4$-dimensional theory is coupled to fundamental matter the symmetry is extended to $\sG_2\times\sU(2r)_{\rm fla}$.) This group can be rotated into its maximal torus
\begin{align}
\sT_{\vec t} = \begin{cases}
\sT_{\vec \tta} \, \times \, \sT_{\vec  \varepsilon} \ ,   \quad &n=2,3\\
\sT_{\vec \tta} \, \times \, \sT_{\vec  \varepsilon} \, \times \, \sT_{\vec m} \ , \quad &n=4 \end{cases}
\end{align}
where we label the torus factors by the corresponding complexified Cartan subalgebra elements $\vec t\ $: the Coulomb parameters $\vec \tta=(\tta_1,\dots,\tta_r)$ are vacuum expectation values of the complex Higgs field $\Phi$ and $\vec  \varepsilon=(  \varepsilon_1,\dots,  \varepsilon_n)$ are the weights for the natural complex scaling action of the $n$-torus $(\FC^\times)^n$ on $\FC^n$, while $\vec m = (m_1,\dots,m_r)$ are masses of $r$ fundamental fermion fields.

We restrict the path integral to gauge field configurations which are $\sT_{\vec t\ }$-invariant.
This can be achieved by coupling the gauge theory to Nekrasov’s $\Omega$-background~\cite{SWcounting} through a shift of the BRST operator by inner contraction
with the vector field generating the $\sSO(2n)$ rotational isometries of $\FR^{2n}\simeq\FC^n$, restricted to $\sU(2)$, $\sU(3)$ and $\sSU(4)$ holonomy which preserve the instanton equations on $\mathbb{C}^n$ for $n=2,3,4$, respectively. The $\Omega$-deformation does not change the instanton equations, but it provides a natural compactification of the moduli space $\frM_{k,r}(\FC^n)$ by giving $\FC^n$ a finite equivariant volume \smash{$
\oint_{\,\FC^n} \, 1 = \frac1{e_{\sT_{\vec  \varepsilon}}(T_{\vec 0}\FC^n)} = \frac1{  \varepsilon_1\cdots  \varepsilon_n}$}.
However, it deforms the equations \eqref{eq:ZPhi} for the Higgs field $\Phi$ to
\begin{align}
[\Phi,Z_a]=  \varepsilon_a\,Z_a\, \quad , \quad [\Phi,Z_{\bar a}^\dagger]=-  \varepsilon_a\,Z_{\bar a}^\dagger \ .
\end{align}
For the NS-type instantons \eqref{eq:NStype} in rank~$1$ the solutions are
\begin{align}
\Phi=S_\ell \, \Big(\,\mbox{$\sum\limits_{a=1}^n$} \,   \varepsilon_a \, N_a \Big) \, S_\ell^\dagger + \tta \, \ident_\CH \ .
\end{align}

The $\Omega$-deformation localizes integrals over the virtual fundamental cycle $[\frM_{k,r}(\FC^n)]^{\rm vir}$ onto the isolated $\sT_{\vec t\ }$-fixed points of the moduli space, and the equivariant integral \smash{$\oint_{\,[\frM_{k,r}(\FC^n)]^{\rm vir}} \,e(\CCK_{k,r})$} is the pushforward of $e(\CCK_{k,r})$ to a point in the $\sT_{\vec t\ }$-equivariant cohomology of $\frM_{k,r}(\FC^n)$, whose coefficient ring is \smash{$\FC[\vec t \ ]$}. The virtual localization theorem~\cite{Graber} then computes the integrals in \eqref{eq:pf} as a sum over these fixed points, giving the instanton partition function as a formal power series in $ q$ and the equivariant parameters $\vec t\, $:
\begin{align}\label{eq:ZCn}
& \!\!\!\!\! Z_{\FC^n}^r( q; \! \vec t \ ) := \sum_{k=0}^\infty \, q^k \ \oint_{[\frM_{k,r}(\FC^n)]^{\rm vir}} \  e(\CCK_{k,r}) \\[4pt]
& = \sum_{k=0}^\infty  q^k \! \sum_{\vec p\in\frM_{k,r}(\FC^n)^{\sT_{\vec t}}} \!\!\!\!\!\!\!\!\! \frac{e_{\sT_{\vec t}}\big((\Ob_{k,r})_{\vec p}\big) \, e_{\sT_{\vec t}}\big((\CCK_{k,r})_{\vec p}\big)}{e_{\sT_{\vec t}}\big(T_{\vec p\,}\frM_{k,r}(\FC^n)\big)} \ . \nonumber
\end{align}
We shall now explain how to evaluate the $\sT_{\vec t\ }$-equivariant Euler classes appearing in this expression as bosonic and fermionic fluctuation determinants in the path integral for the noncommutative cohomological gauge theory on the D$(2n)$-branes~\cite{Moore:1997dj}, which represent quantum fluctuations around $\sT_{\vec t\ }$-invariant instanton solutions.

\subsection{Combinatorics}

The fixed points of the instanton moduli spaces have a combinatorial significance. 
Recall from Section~\ref{subsec:Hilbertscheme} that each $\sU(1)$ noncommutative instanton of charge $k$ corresponds to an ideal of codimension $k$ in the polynomial ring $\FC[w_1,\dots, w_n]$. A $\sT_{\vec\varepsilon\, }$-fixed point of the instanton moduli space corresponds to a \emph{monomial} ideal $\CI$, which in turn can be identified with a set of lattice points~\cite{Nekrasov:2003rj,Okounkov:2006wj}
\begin{align}
\!\! \! \mathscr{Y}_\CI =\Big\{ (b_1,\dots,b_n)\in\mathbbm{Z}_{>0}^n \, \Big\vert  \,  \mbox{$\prod\limits_{a=1}^n\,w_a^{b_a-1}$} \notin \CI \Big\} \ . 
\end{align}

\begin{figure}[h]%
\centering
\includegraphics[width=0.15\textwidth,height=0.1\textheight]{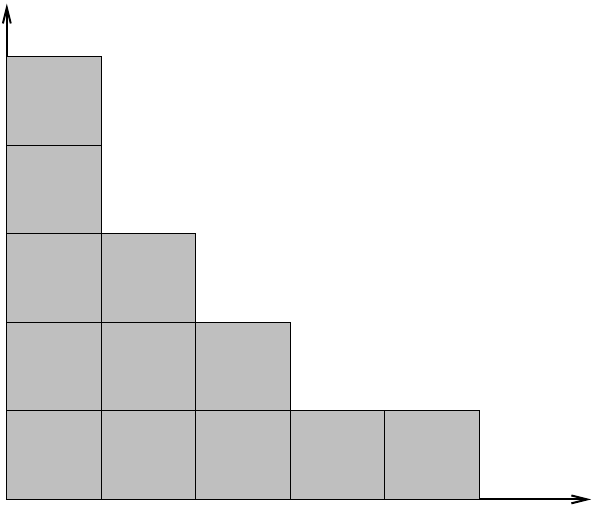} \qquad \qquad
\includegraphics[width=0.15\textwidth,height=0.1\textheight]{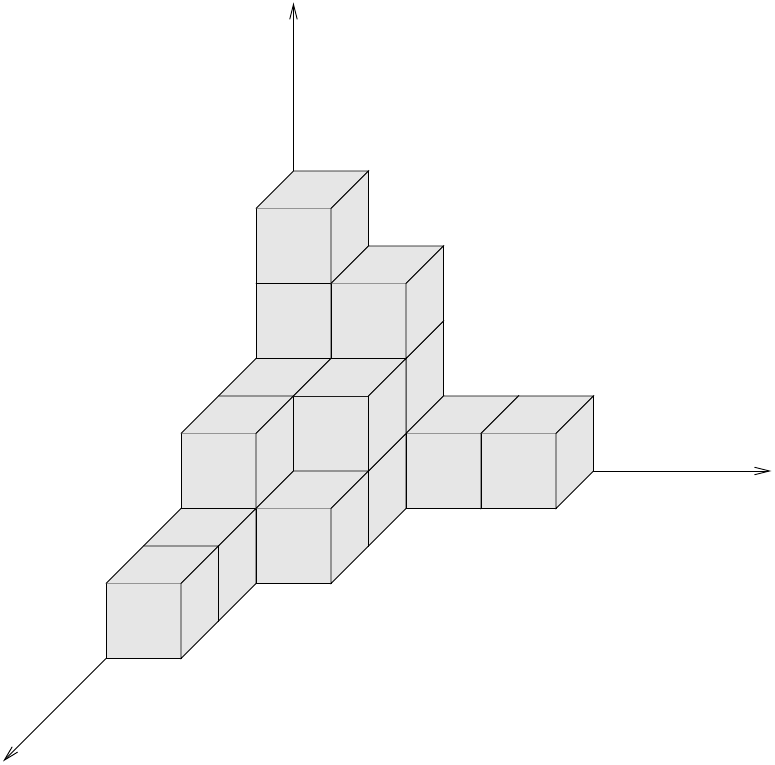}
\caption{Left: A partition represented as a Young diagram with $12$ boxes. Right: A plane partition represented as a $3$-dimensional Young diagram with $18$ cubes.
}\label{fig:Young}
\end{figure}

These can be regarded as labelling the corners of $k$ unit hypercubes piled on top of each other in the positive $n$-multiant of $\FR^n$, called an $n$-dimensional Young diagram, generalizing the more familiar (ordinary) Young diagrams which are obtained for $n=2$. They correspond to $n$-dimensional partitions, i.e. non-increasing sequences of non-negative integers $p=\{p_{i_1\cdots i_{n-1}}\}_{i_1,\dots,i_{n-1}\geq1}$: the $n$-dimensional Young diagram $\CCY$ is recovered from $p$ as $\CCY=\{(b_1,\dots,b_n)\in\RZ^n_{>0}~\vert ~1\leq b_n\leq p_{b_1\cdots b_{n-1}}\}$, where the number of hypercubes in $\CCY$ is the size \smash{$k=\vert  \CCY\vert :=\sum_{i_1,\dots,i_{n-1}\geq1}\, p_{i_1\cdots i_{n-1}}$}. For $n=2$ these are the familiar partitions (see e.g.~\cite{Nakajima:2003pg,Szabo:2015wua}), while for $n=3$ they are \textit{plane} partitions (see e.g.~\cite{Iqbal:2003ds,Cirafici:2008sn,Szabo:2015wua}) and for $n=4$ \textit{solid} partitions (see e.g.~\cite{Nekrasov:2017cih,inprep}). In Figure~\ref{fig:Young} we illustrate the cases $n=2$ and $n=3$.

A $\sT_{\vec t\ }$-invariant $\sU(r)$ noncommutative instanton of charge $k$ corresponds to a decomposition $\CI=\bigoplus_{l=1}^r\,\CI_l$, where $\CI_l$ are monomial ideals of codimension $k_l$ in the polynomial ring $\FC[w_1,\dots, w_n]$ with $\sum_{l=1}^r\,k_l=k$. These are in one-to-one correspondence coloured $n$-dimensional Young diagrams $\vec\CCY=(\CCY_1,\dots,\CCY_r)$, where $\CCY_l$ has $k_l$ hypercubes, or equivalently with coloured $n$-dimensional partitions $\vec p=(p_1,\dots,p_r)$, where~$\vert  \CCY_l\vert  = k_l$. 

From the string theory perspective, the $r$ D$(2n)$-branes are separated in the Coulomb phase of the gauge theory, and the Coulomb parameters label the positions of neighbouring D$(2n)$-branes relative to one another. The hypercubes of the $n$-dimensional Young diagrams index the D$0$-branes, while their colour is the information relative to which D$(2n)$-brane they are bound to: $k_l$ D$0$-branes
bound to a D$(2n)$-brane labelled by $\tta_l$ are described by an $n$-dimensional Young diagram with $k_l$ boxes in the $l$-th sector of the Hilbert space $\CH_{\CI_l}$. They correspond to a charge $k_l$ noncommutative $\sU(1)$ instanton on the worldvolume of the D$(2n)$-brane in position $\tta_l$.

\subsection{Characters}

All ingredients needed for the evaluation of the equivariant instanton partition functions \eqref{eq:ZCn} for the cohomological gauge theory on $\FC^n$ can be encoded into a normalized character $\Ch_\CI(t)$~\cite{Moore:1997dj}, which is a function of a real variable $t\in\FR_{\geq0}$. For the contribution of a noncommutative instanton associated to a collection of ideals $\CI$, it is defined using traces of powers of the Higgs field $\Phi$ as
\begin{align}
\Ch_{\CI}(t) :\!&=\mbox{$\prod\limits_{a=1}^n$}\,(1-\ee^{\,t\,\varepsilon_a}) \ \Tr_{\CH_{\CI}}\,\ee^{\,t\,\Phi}\nonumber \\[4pt]
& =\mbox{$\sum\limits_{l=1}^r$} \, \ee^{\,t\,\tta_l}\, \Big(1- \mbox{$\prod\limits_{a=1}^n$}\,(1-\ee^{\,t\,\varepsilon_a}) \label{eq:character}\\
& \hspace{1cm} \times \mbox{$\sum\limits_{(b_1,\dots,b_n)\in\mathscr{Y}_l}$} \, \ee^{\,t\,\sum_{a=1}^n\,\varepsilon_a\,(b_a-1)}\Big) \ . \nonumber
\end{align}

This computes the top equivariant Chern class of the gauge bundle at a BPS state $\vec\CCY$ through
\begin{align}
{\rm ch}^{\sT_{\vec t}}_n(\CE_\CI) = \oint_{\FC^n}\,\Ch^{(n)}_\CI = \frac{\Ch^{(n)}_\CI}{\varepsilon_1\cdots\varepsilon_n} \ ,
\end{align}
where 
\begin{align}
\!\!\!\! \Ch_\CI ^{(n)} = \mbox{$\sum\limits_{l=1}^r$}\Big(\frac{\tta_l^n}{n!} {-} (-1)^n\, \varepsilon_1\cdots\varepsilon_n\!\!\mbox{$\sum\limits_{(b_1,\dots,b_n)\in\mathscr{Y}_l}$} 1\Big)
\end{align}
is the coefficient of $t^n$ in the power series expansion of the character \eqref{eq:character} about $t=0$. The first term gives the classical contribution $\prod_{l=1}^r\,q^{-\tta_l^n/n!\,\varepsilon_1\cdots\varepsilon_n}$ to the path integral (which we drop), while the second term yields the correct weight factor \smash{$q^{\vert \vec{\mathscr{Y}}\vert }$} in \eqref{eq:ZCn} for the sector of topological charge $k=\vert \vec{\mathscr{Y}}\vert  = \sum_{l=1}^r\,\vert \mathscr{Y}_l\vert $.

\subsection{$n{=}2$: Donaldson-Witten Theory}\label{subsec:VWtheory}

Recall that for $n=2$ the obstruction bundle is trivial and the Vafa-Witten partition function is defined by taking the matter bundle $\CCK_{k,r}=T\frM_{k,r}(\FC^2)$, giving the generating function for the Euler characteristics of the instanton moduli spaces $\frM_{k,r}(\FC^2)$, or alternatively the number of bound states of $k$ D$0$-branes inside $r$ D$4$-branes wrapping $M_2=\FC^2$. In this case the ratio of fluctuation determinants in \eqref{eq:ZCn} is trivially equal to~$1$, and the partition function is independent of the  equivariant parameters $(\vec \tta,\vec   \varepsilon\,)$:
\begin{align}\label{eq:ZC2}
 Z_{\FC^2}^r ( q)=\mbox{$\sum\limits_{\vec \CCY}$} \ q^{\vert \vec \CCY\vert }=\hat\eta(q)^{-r} \ ,
 \end{align}
where the inverse of the Euler function 
\begin{align}
\hat\eta(q) = \mbox{$\prod\limits_{n=1}^\infty$} \, (1-q^n)
\end{align}
is the generating function for the number of partitions of fixed size. For $r=1$ this reproduces the known Euler characteristics of the Hilbert schemes $\frM_{k,1}(\FC^2)\simeq\Hilb^k(\FC^2)$~\cite{Nakajimabook}.

The equivariant gauge theory also allows for the definition of the instanton partition functions for a larger class of $\CN=2$ supersymmetric gauge theories in $4$ dimensions, which are called Nekrasov partition functions~\cite{SWcounting,Nekrasov:2003rj}, whereby the rank of the matter bundle $\CCK_{k,r}$ is no longer necessarily equal to the dimension of the instanton moduli space $\frM_{k,r}(\FC^2)$. When ${\rm rk}\,\CCK_{k,r}=\dim\frM_{k,r}(\FC^2)$ the field theory is conformal, whereas when ${\rm rk}\,\CCK_{k,r}<\dim\frM_{k,r}(\FC^2)$ it is asymptotically free. 

An important example is the instanton partition function of the cohomological gauge theory without matter, known as (equivariant) Donaldson-Witten theory. 
This corresponds to the Nekrasov partition function for the $4$-dimensional $\sU(r)$ pure $\mathcal{N}=2$ gauge theory, defined by the equivariant integral of $1$ over $\frM_{k,r}(\FC^2)$. In this case the Atiyah-Bott localization theorem gives
\begin{align}\label{eq:pfpure2d}
\begin{split}
\!\!\! Z_{\FC^2}^r( \Lambda; \!\! \vec \tta,\vec \varepsilon \ )^{\textrm{pure}} = \sum_{\vec \CCY}\,\frac{\Lambda^{\vert \vec\CCY\vert }}{e_{\sT_{\vec t} }\big(T_{\vec\CCY\,}\frM_{\vert \vec\CCY\vert ,r}(\FC^2)\big)}
 \end{split}
\end{align}
where $\Lambda$ is the UV scale. The function $\varepsilon_1\,\varepsilon_2 \log Z_{\FC^2}^r( \Lambda; \!\! \vec \tta,\vec \varepsilon \,)^{\textrm{pure}}$ is analytic in $(\varepsilon_1,\varepsilon_2)$ near $\varepsilon_1=\varepsilon_2=0$ and its leading term is the prepotential for the Seiberg-Witten effective action~\cite{Seiberg:1994rs}. Geometrically, \eqref{eq:pfpure2d} is the generating function for the equivariant volumes of the instanton moduli spaces $\frM_{k,r}(\FC^2)$.

The equivariant Euler classes can be calculated with the result
\begin{align}
& e_{\sT_{\vec t}}\big(T_{\vec\CCY\,}\frM_{k,r}(\FC^2)\big) \nonumber \\[4pt]
&  ={\det}_{\CH_\CI}(\ad\,\Phi+\varepsilon_1\,\ident_{\CH_\CI}) \, {\det}_{\CH_\CI}(\ad\,\Phi+\varepsilon_2\,\ident_{\CH_\CI}) \nonumber \\[4pt]
&  = \exp\Big(\!\!-\frac14\int_0^\infty\frac{\dd t}{t}\frac{(\ee^{\,t\,\varepsilon_1} {+} \ee^{\,t\,\varepsilon_2})\,\Ch_\CI(t)\,\Ch_\CI(-t)}{(1-\cosh t\,\varepsilon_1)(1-\cosh t\,\varepsilon_2)}\Big) \nonumber\\[4pt]
&  = \mbox{$\prod\limits_{l,l'=1}^r$} \, {\tt N}^{\vec\CCY}_ {l,l'}(\vec\tta,\vec\varepsilon\,) \ ,
\end{align}
where we used a trace regularization for the infinite-dimensional determinants. We set
\begin{align}
& {\tt N}^{\vec\CCY}_ {l,l'}(\vec\tta,\vec\varepsilon\,) \\[4pt]
& :=\mbox{$\prod\limits_{\textrm{\tiny$\square$}\in\CCY_l}$} \big(\tta_{l'l} - {\tt L}_{\CCY_{l'}}(\textrm{\tiny$\square$})\,\varepsilon_1 + \big({\tt A}_{\CCY_l}(\textrm{\tiny$\square$})\,\varepsilon_2 + 1\big)\big) \nonumber \\
&\quad\ \times \mbox{$\prod\limits_{\textrm{\tiny$\square$}'\in\CCY_{l'}}$} \big(\tta_{l'l} + \big({\tt L}_{\CCY_{l}}(\textrm{\tiny$\square$}')+1\big)\,\varepsilon_1 +  {\tt A}_{\CCY_{l'}}(\textrm{\tiny$\square$}')\,\varepsilon_2\big) \nonumber
\end{align}
with $\tta_{l'l}=\tta_{l'}-\tta_{l}$, where ${\tt L}_\CCY(\textrm{\tiny$\square$})$ is the leg-length and ${\tt A}_\CCY(\textrm{\tiny$\square$})$ the arm-length of the box $\textrm{\tiny$\square$}$ in the Young diagram $\CCY$ (see e.g.~\cite{Nakajima:2003uh, Nakajima:2003pg}). 

In particular, the $\sU(1)$ Nekrasov partition function is independent of the Coulomb parameter $\tta$ and the combinatorial series can be summed in closed form using topological properties of Hilbert schemes~\cite{Nakajima:2003pg}, with the result
\begin{align}
 Z_{\FC^2}^{r=1}( \Lambda;\!\!\varepsilon )^{\textrm{pure}}=
 \ee^{\,\Lambda/\varepsilon_1\,\varepsilon_2} \ ,
\end{align} 
which gives $\oint_{\,\Hilb^k(\FC^2)}\,1 = \big(\frac1{\varepsilon_1\,\varepsilon_2}\big)^k$.

\subsection{$n{=}3$: Donaldson-Thomas Theory}\label{subsec:DT3theory}

For the noncommutative gauge theory on $\FC^3$, the ratio of fluctuation determinants required in \eqref{eq:ZCn} is given by \cite{Iqbal:2003ds,Cirafici:2008sn}
\begin{align}\label{eq:fluctationd3d}
&\frac{e_{\sT_{\vec t}}\big((\Ob_{k,r})_{\vec \CCY}\big)}{e_{\sT_{\vec t}}\big(T_{\vec\CCY\,}\frM_{k,r}(\FC^3)\big)} \\[4pt] &=\frac{{\det}_{\CH_\CI}(\ad\,\Phi)\,\prod\limits_{a<b}\,{\det}_{\CH_\CI}(\ad\,\Phi+\varepsilon_{ab}\,\ident_{\CH_\CI})}{{\det}_{\CH_\CI}(\ad\,\Phi+\varepsilon\,\ident_{\CH_\CI}) \! \prod\limits_{a=1}^3\!{\det}_{\CH_\CI}(\ad\,\Phi+\varepsilon_a\ident_{\CH_\CI})} \nonumber \\[4pt]
& = \exp\Big(\!\!-\int_0^\infty \, \frac{\dd t}{t} \, \frac{\Ch_\CI (t)\,\Ch_\CI(-t)\,\ee^{\,t\,\varepsilon}}{\prod_{a=1}^3\,(1-\ee^{\,t\,\varepsilon_a})}\Big) \nonumber
\end{align}
where we defined $\varepsilon_{ab}:=\varepsilon_a+\varepsilon_b$ for $a,b\in\{1,2,3\}$, and $\varepsilon:=\varepsilon_1+\varepsilon_2+\varepsilon_3$. The evaluation of the integral in \eqref{eq:fluctationd3d} at a generic point $\vec\varepsilon$ of the $\Omega$-background is performed in~\cite{Szabo:2015wua}.

Restricting the holonomy group to $\sSU(3)\subset\sU(3)$ sets the 
Calabi-Yau specialization $\varepsilon=0$  of the $\Omega$-deformation. In this case the ratio of fluctuation determinants \eqref{eq:fluctationd3d} simplifies drastically, and the equivariant Euler classes of the tangent and obstruction bundles at a BPS state $\vec\CCY$ coincide up to the sign $(-1)^{r\,\vert \vec\CCY\vert }$, independently of the equivariant parameters $(\vec\tta,\vec\varepsilon\,)$~\cite{Iqbal:2003ds,Cirafici:2008sn,Szabo:2015wua}. The partition function \eqref{eq:ZCn} for $n=3$ then reads as
\begin{align}\label{eq:pf3CY}
\begin{split}
 Z_{\FC^3}^{r} (q)\big\vert _{\varepsilon=0} &= \mbox{$\sum\limits_{\vec \CCY}$} \ (-1)^{r\,\vert \vec \CCY \vert }\, q^{\vert \vec \CCY \vert } \\[4pt]
&= M\big((-1)^r\,q\big)^r \ ,
 \end{split}
\end{align}
where the MacMahon function
\begin{align}
M(q)=\mbox{$\prod\limits_{n=1}^\infty$} \, \frac{1}{(1-q^n)^n}
\end{align}
is the generating function for the number of plane partitions of fixed size.

For $r=1$ the gauge theory partition function \eqref{eq:pf3CY} is the generating function for the Donaldson-Thomas invariants of the Calabi-Yau $3$-fold $M_3=\FC^3$, which enumerate the virtual numbers of ideal sheaves in the Hilbert scheme $\frM_{k,1}(\FC^3)\simeq\Hilb^k(\FC^3)$. It coincides with the partition function of topological string theory on $M_3$ through the simple change of variables $q=-\ee^{-g_s}$~\cite{Iqbal:2003ds,Maulik:2003rzb}, whose formal power series expansion in the string coupling $g_s$ computes the Gromov-Witten invariants which enumerate holomorphic curves in $M_3$. 

For $r>1$ the partition function \eqref{eq:pf3CY} generates higher rank Coulomb branch invariants, which can be obtained as a degenerate central charge limit of higher rank Donaldson-Thomas invariants for D$0$--D$6$ bound states~\cite{Stoppa:2012sf}. These enumerate the virtual numbers of rank $r$ torsion free sheaves which are locally free in codimension~$3$ on the Calabi-Yau $3$-fold $M_3$, and they are also in correspondence with Gromov-Witten invariants. However, the gauge theory in this branch does not seem to be dual to topological string theory. Instead, its physical significance is captured by the generalized Donaldson-Thomas invariants ${\tt DT}_n(M_3)\in\mathbbm{Q}$, which are independent of the rank $r$ of the gauge theory~\cite{Cirafici:2010bd,Cirafici:2011cd} and defined via
\begin{align}
\begin{split}
& Z_{\FC^3}^r(q)\big\vert _{\varepsilon=0} \\[4pt]
& \quad = \exp\Big(-\mbox{$\sum\limits_{n=1}^\infty$} \, (-1)^{n\,r}\,n\,r\ \texttt{DT}_n(\FC^3)\, q^n\Big) \ .
\end{split}
\end{align} 
These are related to the generalized Gopakumar-Vafa BPS invariants ${\tt BPS}_n(M_3)\in\mathbbm{Q}$, which count M2-brane-antibrane bound states in M-theory compactified on $\FR^{1,3}\times M_3\times{\rm S}^1$~\cite{Gopakumar:1998jq} and are defined through $
\texttt{DT}_n(M_3) = \sum_{d\vert  n}\, \texttt{BPS}_{n/d}(M_3)/d^2$. Explicitly one finds
\begin{align}
\texttt{DT}_n(\FC^3)=\sum_{d\vert  n}\,\frac{1}{d^2} \quad , \quad \texttt{BPS}_n(\FC^3)=1 \ .
\end{align}

For general $\sU(3)$ holonomy, with arbitrary $\varepsilon$, it was conjectured in~\cite{Szabo:2015wua} (by comparing with earlier K-theory conjectures of~\cite{Nekrasov:2008kza,Awata:2009dd}) that the equivariant instanton partition function of the $\sU(r)$ cohomological gauge theory on $\FC^3$ is given by
\begin{align}\label{eq:pf3}
 Z_{\FC^3}^{r} (q;\!\vec\varepsilon\,)=M\big((-1)^r\,q\big)^{-\frac{r\,\varepsilon_{12}\,\varepsilon_{23}\,\varepsilon_{13}}{\varepsilon_1\,\varepsilon_2\,\varepsilon_3}} \ ,
\end{align}
independently of the Coulomb parameters $\vec\tta$.
For $r=1$ this formula appeared originally in~\cite{Maulik:2004txy} in the geometric setting of relative Donaldson-Thomas theory, where the exponent is minus the $\sT_{\vec\varepsilon\,}$-equivariant Euler characteristic $\chi_{\sT_{\vec\varepsilon\,}}(M_3) = \oint_{M_3}\,\ch_3^{\sT_{\vec\varepsilon}\,}(TM_3)$ of the K\"ahler $3$-fold $M_3=\FC^3$. For $r>1$ it was proven in~\cite{Fasola:2020hqa} using the symmetric perfect obstruction theory of the Quot scheme $\frM_{k,r}(\FC^3)\simeq\Quot_{\FC^3}^k(\CO^{\oplus r})$.

\subsection{$n{=}4$: Donaldson-Thomas Theory}\label{subsec:DT4theory}

For the noncommutative gauge theory on $\FC^4$, the ratio of fluctuation determinants associated to the tangent-obstruction theory is given by~\cite{inprep}
\begin{align}
&\!\!\! \frac{e_{\sT_{\vec t}}\big((\Ob_{k,r}^-)_{\vec \CCY}\big)}{e_{\sT_{\vec t}}\big(T_{\vec\CCY\,}\frM_{k,r}(\FC^4)\big)} \label{eq:flu4d} \\[4pt] & \hspace{.5cm} =\frac{{\det}_{\CH_\CI}(\ad\,\Phi)\prod\limits_{\stackrel{\scriptstyle a,b=1}{\scriptstyle a<b}}^3 {\det}_{\CH_\CI}(\ad\,\Phi+\varepsilon_{ab}\,\ident_{\CH_\CI})}{\prod\limits_{a=1}^4\, {\det}_{\CH_\CI}(\ad\,\Phi+\varepsilon_a\,\ident_{\CH_\CI})} \nonumber \\[4pt]
& \hspace{.5cm} = \exp\Big(\!\!-\frac12\,\int_0^\infty\,\frac{\dd t}{t}\,\frac{\Ch_\CI (t)\,\Ch_\CI(-t)}{\prod_{a=1}^4\,(1-\ee^{\,t\,\varepsilon_a})}\Big) \ . \nonumber
\end{align}
This expression has a sign ambiguity due to the choice of square root needed to define the equivariant Euler class of the real self-dual obstruction bundle $\Ob_{k,r}^-$, which is related to its orientation. The $\sSU(4)$ holonomy implies that the equivariant parameters $\vec\varepsilon=(\varepsilon_1,\varepsilon_2,\varepsilon_3,\varepsilon_4)$ satisfy the Calabi-Yau constraint $\varepsilon_1+\cdots+\varepsilon_4=0$, which enables us to express $e_{\sT_{\vec t}}\big((\Ob^-_{k,r})_{\vec \CCY}\big)$ using any triple $(\varepsilon_a,\varepsilon_b,\varepsilon_c)$, up to a sign; in writing \eqref{eq:flu4d} we have chosen the triple $(\varepsilon_1,\varepsilon_2,\varepsilon_3)$.

The contribution to \eqref{eq:ZCn} for $n=4$ from the matter bundle $\CCK_{k,r} = \CCV_{k,r}\otimes\FC^r$ is likewise given by~\cite{inprep}
\begin{align}
& e_{\sT_{\vec t}}\big((\CCV_{k,r})_{\vec\CCY}\otimes\FC^r\big)=\mbox{$\prod\limits_{l=1}^r$} \, {\det}_{\CH_\CI}(\Phi-m_l\,\ident_{\CH_\CI}) \nonumber \\[4pt]
& \ \ = \exp\Big(\!\!-\int_0^\infty\,\frac{\dd t}{t}\,\frac{\sum_{l=1}^r\,\Ch_\CI (t)\,\ee^{-t\,m_l}}{\prod_{a=1}^4\,(1-\ee^{t\,\varepsilon_a})} \Big) \ . \label{eq:intful4d}
\end{align}

The evaluation of the integrals in \eqref{eq:flu4d} and \eqref{eq:intful4d} is performed in \cite{inprep}, where it is shown that the $8$-dimensional cohomological partition function can be summed to the closed formula
\begin{align}
Z_{\FC^4}^r(q;\!\!\vec{\varepsilon},m)=M(-q)^{-\frac{m\,r\,\varepsilon_{12}\,\varepsilon_{13}\,\varepsilon_{23}}{\varepsilon_1\,\varepsilon_2\,\varepsilon_3\,\varepsilon_4}} \ , \label{PfRank}
\end{align}
where $m=\frac{1}{r} \, \sum_{l=1}^r\,(m_l-\tta_l)$. This formula demonstrates an intimate relation between the instanton counting theories in
$8$ and $6$~dimensions. While this connection is clear from the field theory perspective~\cite{Nekrasov:2017cih,inprep}, it is non-trivial from a combinatorial
point of view: the partition function \eqref{eq:ZCn} for $n=4$ is written as a sum over (coloured) \emph{solid} partitions, for which no generating function is known and instead the generating function of \emph{plane} partitions appears. 

In particular, by setting $m_l=\tta_l+\varepsilon_4$ for $l=1,\dots,r$, the partition function \eqref{PfRank} reduces to the partition function \eqref{eq:pf3} for the cohomological gauge theory on the K\"ahler $3$-fold $M_3=\FC^3$, up to the rescaling $q\to(-1)^{r+1}\,q$. It would be interesting to understand this mass specialization as a specific charge configuration of the D$8$-branes and anti-D$8$-branes that annihilate into D$6$-branes through a process of tachyon condensation~\cite{Sen:1998sm}, whose
bound states with the D$0$-branes correspond to $6$-dimensional instantons.

For $r=1$ the formula \eqref{PfRank} also appears in~\cite{Nekrasov:2017cih} as the cohomological limit of a conjectural formula for the analogue K-theory partition function on $\FC^4\times{\rm S}^1$. In the geometric setting of~\cite{Cao:2017swr}, the exponent is the $\sT_{\vec\varepsilon\,}$-equivariant characteristic number $\oint_{M_4}\,\ch^{\sT_{\vec\varepsilon}}_1(L) \wedge\ch_3^{\sT_{\vec\varepsilon}}(TM_4)$, where $L\to M_4$ is a line bundle of Chern number $-m$ which represents the Chan-Paton gauge bundle on the anti-D$8$-brane; the specialisation $m=\varepsilon_4$ corresponds to the $\sT_{\vec\varepsilon\,}$-equivariant line bundle $L=\CO(D)$ associated to the divisor $D=\{\vec z\in\FC^4\,\vert \,z_4=0\}\simeq\FC^3$. Unlike the theory on Calabi-Yau $3$-folds, here the rank~$1$ gauge theory does not seem to be dual to a Gromov-Witten theory for Calabi-Yau $4$-folds (see e.g.~\cite{Klemm:2007in}). For $r>1$ the formula \eqref{PfRank} also follows from the cohomological limit of the conjectural higher rank K-theory formula of~\cite{Nekrasov:2018xsb} (see~\cite{inprep}). 

As in Section~\ref{subsec:VWtheory}, the $\Omega$-deformation also allows for the definition of an instanton partition function by dropping the matter bundle contribution. Then the equivariant instanton partition function of pure $\sU(r)$ supersymmetric Yang-Mills theory in $8$ dimensions is given by~\cite{inprep}
\begin{align}
Z_{\FC^4}^{r}(\Lambda;\!\!\vec\varepsilon\,)^{\rm pure}= \begin{cases}
\ee^{-\Lambda\,\frac{\varepsilon_{12}\,\varepsilon_{13}\,\varepsilon_{23}}{\varepsilon_1\,\varepsilon_2\,\varepsilon_3\,\varepsilon_4}} \ , \  & r=1 \\
1 \ , \  & r>1 
\end{cases} \ .
\end{align} 
This follows from the large mass limit $m_l\to\infty$ of \eqref{PfRank}, with $q\to0$ and $\Lambda=(-1)^r\,m_1\cdots m_r\,q$, which decouples the fundamental hypermultiplets.
In the rank~$1$ case, this formula was conjectured in~\cite{Nekrasov:2017cih,Cao:2017swr}; geometrically it implies that \smash{$\oint_{\,[\Hilb^k(\FC^4)]_{\mathfrak o}^{\rm vir}}\,1 = \big(-\frac{\varepsilon_{12}\,\varepsilon_{13}\,\varepsilon_{23}}{\varepsilon_1\,\varepsilon_2\,\varepsilon_3\,\varepsilon_4}\big)^k$}  (with our choice of orientation $\mathfrak o$). However, in contrast to the pure $4$-dimensional gauge theory \eqref{eq:pfpure2d}, here the higher rank pure $\CN=2$ partition functions on $\FC^4$ are trivial: the equivariant volumes of the instanton moduli spaces $\frM_{k,r}(\FC^4)\simeq\Quot_{\FC^4}^k(\CO^{\oplus r})$ for $r>1$ all vanish.

\section{Orbifolds, Defects \\ and Toric Geometry}\label{sec:further}

\subsection{Instantons on Orbifolds}\label{subsec:orbifold}

Type II string compactifications on singular spaces such as orbifolds can be well-defined, and we may use these as a first non-trivial generalization of our constructions beyond the flat spaces $\FC^n$. A central role is played by orbifolds that admit crepant resolutions, which preserve Calabi-Yau properties and allow computation of the string spectrum on a smooth Calabi-Yau space obtained by blow-ups of the orbifold singularities~\cite{Alexandrov:2011va}. The string dynamics can then be considered in `orbifold' or `large radius' phases, which are related by collapsing the compact cycles of the resolution~\cite{Douglas:2000qw}. 

This enables the extension of the noncommutative instanton theories to the toric Calabi-Yau orbifolds $M_n=\mathbbm{C}^n/\sfGamma$, where $\sfGamma\subset\sSL(n,\mathbbm{C})$ is a finite abelian group which commutes with the torus $\sT_{\vec\varepsilon}$ of the $\Omega$-background, and $\FC^n$ is viewed as the fundamental representation of $\sSL(n,\FC)$. As a $\sGamma$-module, $\FC^n$ has branching weights $\vec s=(s_1,\dots,s_n)$. We denote by \smash{$\widehat \sfGamma$} the finite abelian group of irreducible representations $\rho_s$ of~$\sGamma$.

The orbifold field theory is naturally constructed by keeping only contributions from $\sGamma$-invariant instantons, via projection to the trivial representation $s=0$ of the orbifold group $\sGamma$, which involves an action of $\sGamma$ on the Chan-Paton space $\FC^r$, defined by a homomorphism $\sGamma\to\sU(r)$. Focusing momentarily on the rank~$1$ case, the Hilbert space \eqref{eq:Fock} for the noncommutative gauge theory is a $\sGamma$-module which admits an isotopical decomposition into irreducible representations 
\begin{align}
\CH=\textstyle\bigoplus\limits_{s\in \widehat \Gamma}\, \CH_s \ , 
\end{align}
where $\CH_s$ is the subspace of $\CH$ spanned by states $\vert\vec n\,\rangle$ satisfying $\vec n\cdot\vec s := \sum_{a=1}^n\,n_a\,s_a\equiv s$, by which is meant $\bigotimes_{a=1}^n\,\rho_{s_a}^{\otimes n_a}\simeq\rho_{s}$ at the level of $\sGamma$-modules. 

The $\sGamma$-equivariant operators $\vec Z:\CH\to\FC^n\otimes\CH$ decompose  into maps \smash{$Z_a^{(s)}:\CH_s\to \CH_{s+s_a}$}, and the noncommutative instanton equations \eqref{alg_insteqd8} become
\small
\begin{align}\label{alg_insteqd8orb}
& \!\!\!\! Z_a^{(s+s_b)}\,Z_b^{(s)} -Z_b^{(s+s_a)}\,Z_a^{(s)} \\[4pt]
&= \begin{cases}
0 \ , \ n=2,3 \\ \tfrac{1}{2}\,\epsilon_{ab\bar c\bar d}\,\big(Z_{\Bar{c}}^{\dagger (s-s_d-s_c)}\,Z_{\Bar{d}}^{\dagger (s-s_d)} \\
\qquad\qquad - Z_{\Bar{d}}^{\dagger (s-s_d-s_c)}\,Z_{\Bar{c}}^{\dagger (s-s_c)}\big)\ , \ n=4 \ ,
\end{cases} \nonumber \\[4pt]
&\textstyle\sum\limits_{a=1}^n \, \big(Z_{\Bar{a}}^{(s)\dagger }\, Z_a^{(s)} - Z_{\Bar{a}}^{ (s-s_a)}\, Z_a^{ (s-s_a)\dagger}\big) = -n \, \xi \, P^{(s)} \ , \nonumber
\end{align}
\normalsize
where $P^{(s)}$ is  the projection of $\CH$ onto the isotopical subspace $\CH_s$, and \eqref{alg_insteqd8orb} should be understood with the Calabi-Yau condition $\vert\vec s\,\vert\equiv0$. Similarly, the $\sGamma$-equivariant Higgs field \smash{$\Phi:\CH\to\CH$} splits into maps $\Phi^{(s)}:\CH_s\to\CH_s$ obeying the equations \smash{$\Phi^{(s+s_a)}\,Z_a^{(s)} - Z_a^{(s)}\,\Phi^{(s)} = \varepsilon_a\,Z_a^{(s)}$}.

Partial isometries $T_k$ of $\CH$ decompose accordingly, and in the equivariant gauge theory the solutions of \eqref{alg_insteqd8orb} are parametrized by  $\widehat \Gamma$-coloured $n$-dimensional Young diagrams \smash{$\mathscr{Y}=(\mathscr{Y}_s)_{s\in\widehat \Gamma}$}, where $\vec b=(b_1,\dots, b_n)\in\mathscr{Y}_s$ if and only if \smash{$\vec s \cdot\vec b\equiv s$}. Every diagram $\mathscr{Y}_s$ can be identified with a monomial ideal $\CI_s$ in the polynomial ring $ \FC[w_1,\dots, w_n]$ of codimension $k_s$ such that $k=\sum_{s\in\widehat\sfGamma}\, k_s$. Geometrically, $\CCY_s$ corresponds to a $\sGamma$-invariant $0$-dimensional subscheme $Z\subset\FC^n$ whose global sections $H^0(\CO_Z)$ form the representation $\rho_s$.

To compute orbifold instanton partition functions, we define an equivariant character generalizing \eqref{eq:character} which gives the contribution of a noncommutative instanton associated to a direct sum $\CI=\bigoplus_{s\in\widehat\sfGamma}\,\CI_s$ of monomial ideals through
\begin{align}\label{chi_orb}
\begin{split}
\Ch_\CI^{\sfGamma}(t):\!&=\tfrac{1}{\Ch_0^\sfGamma (t)} \, \Tr_{\CH_{\CI_0}} \, \ee^{\,t\,\Phi} \\[4pt]
&= 1- \tfrac{1}{\Ch_0^\sfGamma (t)} \, \textstyle\sum\limits_{\vec b\in\CCY_0} \, \ee^{\,t\,\vec\varepsilon\,\cdot\,\vec b} \ ,
\end{split}
\end{align} 
where
\begin{align}
\Ch_0^\sfGamma (t)=\textstyle\sum\limits_{\CCY'} \ \sum\limits_{\vec b'\in\CCY'_0} \, \ee^{\,t\, \vec\varepsilon\,\cdot\,\vec b'} \ .
\end{align}
More generally, one can weigh the $\widehat\sGamma$-coloured noncommutative instantons in the twisted orbifold sectors $s\neq0$ by a larger set of variables, in order to match with the counting of fractional D$0$-branes expected from string theory in the D$(2n)$-brane gauge theory on $\FC^n/\sGamma$. The instanton partition functions have been calculated for various orbifold groups $\sGamma$ and $n=2,3,4$ in e.g.~\cite{Fucito:2004ry,Dijkgraaf:2007fe,Cirafici:2010bd,Bonelli:2011jx,Bonelli:2011kv,Bonelli:2020gku,inprep}.

\begin{figure}[h]%
\centering
\includegraphics[width=0.26\textwidth]{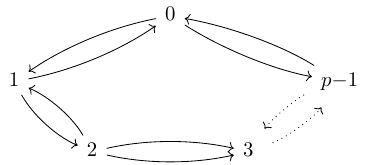} \ 
\includegraphics[width=0.2\textwidth]{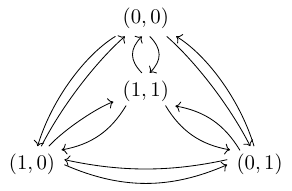}
\caption{McKay quivers of the $A_{p-1}$ orbifold $\FC^2/\mathbbm{Z}_p$ (left) and the orbifold $\FC^3/\mathbbm{Z}_2{\times} \mathbbm{Z}_2$ (right). The vertices are represented by $k_s$ fractional D$0$-branes stuck at the orbifold singularity which transform in the representation $\rho_s$ of $\sGamma$, while the arrows are associated to $\sGamma$-invariant massless excitations of open strings starting and ending on the same D$0$-brane in $\FC^n$. Framing introduces vertices represented by D$(2n)$-branes and arrows represented by D$0$--D$(2n)$ strings as before.
}\label{fig:McKayquiver}
\end{figure}

It is also useful here to consider quiver matrix models associated to the orbifold group $\sfGamma$, generalizing the ADHM quiver description on $\FC^n$. To each representation $\rho_s$ for \smash{$s\in\widehat{\sfGamma}$} we associate a vertex of a quiver, where a vertex $s$ is connected to a vertex $s'$ by a number of arrows equal to the multiplicity of the representation $\rho_{s'}$ in $\FC^n\otimes\rho_s$. The resulting quiver is known as the (generalized) McKay quiver; we illustrate two examples in Figure~\ref{fig:McKayquiver}. It comes with a set of relations which encode the holomorphic parts of the instanton equations \eqref{alg_insteqd8orb}.

This identifies the path algebra of the bounded McKay quiver as a noncommutative resolution of the abelian orbifold singularity $\FC^n/\sGamma$, in the sense that its centre is isomorphic to the (noncommutative) crossed product coordinate algebra $\FC[w_1 ,\dots, w_n ] \rtimes \sGamma$ of the quotient $\FC^n /\sGamma$. The moduli space $\frM_{k,r}^\sGamma(\FC^n)$ of instantons on $\FC^n/\sGamma$, viewed as $\sGamma$-invariant $\sU(r)$ noncommutative instantons on $\FC^n$ of topological charge $k$, is then identified as a moduli space of stable representations of the bounded McKay quiver.

The construction of these orbifold theories depends dramatically on the complex dimension~$n$. They have been studied from this perspective in~\cite{KronNak,Nakajima:1994nid,Douglas:1996sw,Ivanova:2013mea} for $n=2$, in~\cite{Cirafici:2010bd,Cirafici:2011cd} for $n=3$, and in~\cite{Bonelli:2020gku,inprep} for $n=4$. Their connections with (generalized) McKay correspondences for $n=2,3$ is reviewed in~\cite{Cirafici:2012qc}. 

\noindent\fbox{\small{$\boldsymbol{n=2}$}} \ 
The only toric Calabi-Yau orbifolds of $\FC^2$ are the Kleinian $A_{p-1}$ singularities $\FC^2/\RZ_p$, with the generator of the cyclic group $\sGamma=\RZ_p$ acting on $\FC^2$ as $(z_1,z_2)\mapsto (\ee^{\,2\pi\,\ii/p}\,z_1,\ee^{-2\pi\,\ii/p}\,z_2)$. By the McKay correspondence~\cite{Mckay}, the McKay quiver traces out the A-type affine Dynkin diagram of the simply-laced Lie algebra $\mathfrak{su}(p)$ (see~Figure~\ref{fig:McKayquiver}). The connected components of the instanton moduli space \smash{$\frM_{k,r}^{\RZ_p}(\FC^2)$} are called Nakajima quiver varieties of type~A~\cite{Nakajima:1994nid}, which parametrize framed $\RZ_p$-equivariant torsion free sheaves on the projective plane $\PP^2$.

These moduli spaces are realized in type II string theory as Higgs branches of quiver gauge theories which appear as worldvolume field theories on D$p$-branes in a D$p$--D$(p{+}4)$ system with the D$(p{+}4)$-branes located at the fixed point of the orbifold~\cite{Douglas:1996sw}. The D$p$--D$(p{+}4)$ bound states are enumerated by the Vafa-Witten partition function for $\CN=4$ supersymmetric Yang-Mills theory on $\FC^2/\RZ_p$, which reproduces the character of the affine Lie algebra $\widehat{\mathfrak{su}}(p)$~\cite{Nakajima:1994nid}. 

The same result is obtained~\cite{Fujii:2005dk} from the instanton partition function on the minimal resolution of the quotient singularity $\FC^2/\RZ_p$, which is an ALE space $X$ of type $A_{p-1}$. Since $X$ is a toric variety, its equivariant partition function can be computed by gluing together $\FC^2$ patches, as we explain in Section~\ref{subsec:toric}. For general $\CN=2$ theories the two approaches are expected to produce the same partition function, provided that it is independent of the size of the cycles of the resolution $X$. This relation has however not yet been proven; see~\cite{Bruzzo:2013daa,Bruzzo:2014jza} for discussion of this point.

\noindent\fbox{\small{$\boldsymbol{n=3}$}} \ 
Like the case $n=2$, for $n=3$ a crepant resolution $X\to\FC^3/\sGamma$ always exists (though it is no longer unique in general), which enables a parametrization of the instanton  moduli space \smash{$\frM_{k,r}^\sGamma(\FC^3)$} in terms of torsion free sheaves via an application of Beilinson’s theorem~\cite{Cirafici:2010bd}. A natural choice is the $\sGamma$-Hilbert scheme $X=\Hilb^{\sfGamma}(\FC^3)$ of $\sfGamma$-invariant $0$-dimensional subschemes $Z\subset \FC^3$ of length $ \vert\sGamma\vert$ such that $H^0(\mathcal{O}_Z)$ is the regular representation of $\sfGamma$; it can be parametrized by $\sfGamma$-invariant ideals $\CI\subset\FC[w_1,w_2,w_3]$ of codimension $\vert\sGamma\vert$, and it is itself isomorphic to a moduli space of noncommutative instantons in the Coulomb branch. For the example $\sGamma=\RZ_2{\times}\RZ_2$ whose generators act on $\FC^3$ as $(z_1,z_2,z_3)\mapsto(-z_1,-z_2,z_3)$ and $(z_1,z_2,z_3)\mapsto(z_1,-z_2,-z_3)$ (see Figure~\ref{fig:McKayquiver}), the natural crepant resolution $X$ is the closed topological vertex geometry~\cite{Cirafici:2010bd,Cirafici:2012qc}.

The generalized McKay correspondence dictates an equivalence between (stable) representations of the bounded McKay quiver and coherent sheaves on a crepant resolution $X$ of the singularity $\FC^3/\sGamma$~\cite{Bridgeland:2001xf,Alastair:2002yt}. This reformulates the enumeration of noncommutative Donaldson-Thomas invariants as an instanton counting problem~\cite{Cirafici:2010bd}, as the noncommutative orbifold instantons enumerate $\sfGamma$-equivariant sheaves on $\FC^3$. For ideal sheaves they coincide with the orbifold Donaldson-Thomas invariants defined in~\cite{Young:2008hn}. Unlike the case $n=2$, the instanton partition functions on $\FC^3/\sGamma$ and its toric crepant resolution $X$ are related by wall-crossing formulas~\cite{Cirafici:2010bd,Cirafici:2011cd}, which connect the orbifold and large radius phases.

\noindent\fbox{\small{$\boldsymbol{n=4}$}} \ 
While for $n=2,3$ the quotient singularities $\FC^n/\sGamma$ always have crepant resolutions~\cite{Bridgeland:2001xf}, for $n=4$ this is not always guaranteed for an arbitrary finite abelian subgroup $\sfGamma\subset \sSL(4,\mathbbm{C})$~\cite{Sato}; even the $\sGamma$-Hilbert scheme $\Hilb^\sGamma(\FC^n)$ can be singular when $n>3$. Moreover, the path algebra of the bounded McKay quiver does not generally give a noncommutative crepant resolution when $n>3$~\cite{Wemyss:2012ee,vandenbergh}; a counterexample is $\FC^4/\RZ_2$ with $\RZ_2$ acting on $\vec z\in\FC^4$ as the reflection $\vec z\mapsto -\vec z$, which has no crepant resolutions.

For orbifold groups $\sGamma$ such that the Hilbert-Chow morphism $\Hilb^\sGamma(\FC^4)\to\FC^4/\sGamma$ is a crepant resolution, a geometric interpretation of the noncommutative instantons on $\FC^4/\sGamma$ in terms of coherent sheaves is sketched in~\cite{Bonelli:2020gku} by extending the $n=3$ constructions of~\cite{Cirafici:2010bd} which are based on homological algebra. This would enable an analogous reformulation of Donaldson-Thomas theory on these orbifolds, as well as a comparison of the orbifold and large radius phases of the low-energy dynamics of D-branes on $\FC^4/\sGamma$. A detailed analysis of instanton partition functions on Calabi-Yau orbifolds of the $4$-fold $\FC^4$ appears in~\cite{inprep}.

\subsection{Instantons on Divisor Defects}\label{subsec:divdefect}

Instantons on orbifolds are also useful for the description of instantons in gauge theories in the presence of divisor defects, on which gauge fields become singular with fixed monodromy. Near a defect the ambient $n$-fold has the local product form $M_n=D \times C$, where $D$ is a divisor and $C$ is a local fiber of the normal bundle. When restricted to $C$, the gauge connection assumes the form 
\begin{align}\label{eq:Adefect}
    A=-\ii\,{\mathsf\alpha} \, \dd \log w \, + \, \text{non-singular terms} \ ,
\end{align}
where $w$ is a local coordinate on $C$, and $\mathsf\alpha$ is an element in the Cartan subalgebra of the gauge group $\sU(r)$ which parametrizes the singularity of the (full) divisor operator. The curvature $2$-form $F$ of \eqref{eq:Adefect} has a singularity proportional to the de~Rham current which is Poincar\'e dual to the divisor $D$.

The pair $(D,\mathsf{\alpha})$ is called a parabolic structure. It breaks the gauge symmetry along $D$ to the commutant subgroup $\sL_{\mathsf\alpha}\subset\sU(r)$ of $\mathsf\alpha$, called a (minimal) Levi subgroup. The Levi subgroups $\sL$ correspond to parabolic subgroups of $\sGL(r,\FC)$, which can be thought of as stabilizers of flags of vector spaces $0=U_0\subset U_1\subset\cdots\subset U_r=\FC^r$.

For $n=2,3$, the moduli space of instantons in the presence of a defect in $M_n=\FC^n$ is parametrized by framed flags
\begin{align}
\mathcal{F}_0(D)\subset\mathcal{F}_{-r+1}\subset\cdots\subset\mathcal{F}_{-1}\subset\mathcal{F}_0
\end{align}
of rank~$r$ torsion free sheaves on $(\PP^1)^n$ with $\ch_1(\mathcal{F}_l)=l\,D$ and $\ch_n(\mathcal{F}_l)=k_l$. These are called parabolic sheaves~\cite{Negut:2011aa} and the corresponding moduli space is denoted by $\mathfrak{D}_{ \vec k , r}(\FC^n)$, with \smash{$\vec k =(k_0,k_1\dots , k_{r-1})$}.

The moduli spaces \smash{$\mathfrak{D}_{ \vec k , r}(\FC^n)$} are the connected components of the moduli space
\smash{$\frM_{k,r}^\sGamma(\FC^n)$}
of $\sU(r)$ noncommutative orbifold instantons on $\FC^n$ with topological charge \smash{$k=\vert\vec k\,\vert$} for the action of a group $\sGamma$ whose fixed point locus is the divisor $D$ and which preserves the parabolic structures; for a detailed discussion see~\cite{Alday:2010vg,Kanno:2011fw, Cirafici:2013nja}. For example, for a defect operator located on the divisor $D=\{\vec z\in\FC^n\,\vert \,z_n=0\}$, the orbifold group is the cyclic group $\sGamma=\mathbbm{Z}_r$ acting on the target space as $(z_1,\dots,z_{n-1},z_n)\mapsto (z_1,\dots,z_{n-1},\ee^{\,2\pi \, \ii/r}\,z_n)$. The analysis of these noncommutative instantons then follows the orbifold quiver description from Section~\ref{subsec:orbifold}~\cite{Kanno:2011fw, Cirafici:2013nja}.

For $n=4$ these types of instantons have not yet been studied in detail. The general setup of the problem is sketched in~\cite{Cirafici:2013nja}. 

\subsection{Spiked/Tetrahedron Instantons}\label{subsec:spiked}

Another generalization of noncommutative instantons on $\FC^{m}$ for $m=2,3$ involves more general supersymmetric defects in the gauge theory on $\FC^4$, which correspond to configurations of instantons on $\FC^{m}\subset\FC^4$ with different spatial orientations. For $m=2$ these are called spiked instantons~\cite{Nekrasov:2015wsu,Nekrasov:2016qym}, while for $m=3$ they are called tetrahedron instantons~\cite{Pomoni:2021hkn}. The noncommuative gauge theory is realized similarly to Section \ref{subsec:NCinsteq}, now with the Hilbert space
\begin{align}
\textstyle\bigoplus\limits_{A\in\CCC_m} \, \mathscr{H}_A=\mbox{$\bigoplus\limits_{A\in\mathscr{C}_m}$} \, \FC^{r_A}\otimes \CH_{1\cdots m}
\end{align}
for $m=2,3$, where $\CH_{1\cdots m}$ is the $m$-oscillator Fock module \eqref{eq:Fock}, and $\mathscr{C}_m$ is the set of $\binom{4}{m}$ labels of $m$-dimensional coordinate hyperplanes in $\FC^4$. 

We seek operator solutions $Z_a$, $a\in\{1,\dots,4\}$ of the BPS equations \eqref{alg_insteqd8}, with $n=4$ and $r = \sum_{A\in\CCC_m}\,r_A$, for noncommutative gauge fields with gauge group $\timesbig_{A\in\CCC_m}\,\sU(r_A)$ such that $Z_{a}\vert _{\FC^{r_A}\otimes \CH_{1\cdots m}}=0$ whenever $a\notin A$.
These are given by suitable linear combinations of solutions \smash{$(\mathtt{Z}_{A,1}, \dots, \mathtt{Z}_{A,m})_{A\in\mathscr{C}_m}$} to $\sU(r_A)$ noncommutative instanton equations on $\FC^m$, determined by partial isometries $T_{A,k}$ of the Hilbert space $\FC^{r_A}\otimes \CH_{1\cdots m}$. For example, a spiked instanton solution (with $m=2$) is given by
\begin{align}\begin{split}
Z_a&=\textstyle\bigoplus\limits_{\stackrel{\scriptstyle A\in\CCC_2}{a\in A}} \,\mathtt{Z}_{A,h_A(a)} \ ,
\end{split}\end{align} 
where $h_{(ab)}(a)=1,2$ according to whether $a<b$ or $a>b$, respectively.
Similarly, the Higgs fields are given by $\Phi=\bigoplus_{A\in\mathscr{C}_m}\,\Phi_A$ where $\Phi_A$ are solutions to $\sU(r_A)$ noncommutative parallel section equations on $\FC^m$. 

\begin{figure}[h]%
\centering
\includegraphics[width=0.3\textwidth]{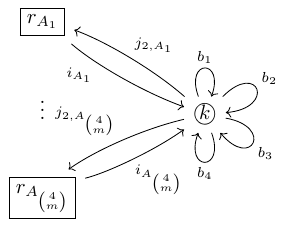}
\caption{ADHM quiver description of spiked (with $j_{2,A}$) and tetrahedron (without $j_{2,A}$) instantons. Vertices labelled by $r_A$ for $A\in\mathscr{C}_m$ are associated to $r_A$ D$(2m{+}1)_A$-branes wrapping the coordinate hyperplane $\FC_A^m\subset\FC^4$, while the remaining vertex is associated to $k$ D$1$-branes. The arrows $i_A,j_{2, A}$ represent the D$1$--D$(2m{+}1)_A$ strings, while $b_1,\dots,b_4$ describe D$1$--D$1$ strings.
}\label{fig:ADHMspiked}
\end{figure}

Spiked/tetrahedron instantons find their physical realizations in type~IIB string theory~\cite{Nekrasov:2016qym,Nekrasov:2016gud,Pomoni:2021hkn} as D$1$-branes probing a configuration of intersecting stacks of D$(2m{+}1)$-branes with an appropriate large constant background $B$-field turned on; the different orientations of the D$(2m{+}1)$-branes are labelled by the hyperplane set $\CCC_m$. The result is an ADHM-type matrix model based on a quiver encoding the field content of the low-energy effective theory on the D$1$-branes (see Figure~\ref{fig:ADHMspiked}).

These solutions parametrize the spiked/tetrahedron instanton moduli space $\frM_{k,\vec r\,}(\FC^m)$, where $\vec r=(r_A)_{A\in\CCC_m}$. If  $\vec r =(r_A,0,\dots,0)$ (up to permutation of its entries), then this is isomorphic to the usual moduli space of noncommutative instantons on $\FC^m$: $\frM_{k,(r_A,0,\dots,0)}(\FC^m)\simeq \Quot^k_{\FC^m}(\CO^{\oplus r_A})$. 

To treat the case of general $\vec r$, consider the worldvolume of $r_A$ D$(2m{+}1)_A$-branes for fixed $A\in\mathscr{C}_m$ which supports $k$ D$1$-branes. These are accompanied by $r_{A^\circ}$ D$(2m{+}1)_{A^\circ}$-branes, labelled by $A^\circ\in\mathscr{C}_m \setminus A$, which intersect the D$(2m{+}1)_A$-branes in hyperplanes 
\begin{align}
\FH_{A^\circ}:=\{\vec z\in\FC_A^m \ \vert  \ z_{a}=0 \, , \, a\notin A^\circ\} \ ,
\end{align}
and produce defects in $\FC_A^m$ of codimension~$1$ or~$2$. Let $Z_A$ denote the union of a set of $k$ points with the collection of hyperplanes $\bigcup_{A^{\circ}\in\CCC_m\setminus A}\,\FH_{A^\circ}$. 

Then each solution realises a quotient \smash{$\CO^{\oplus r_A}\twoheadrightarrow \CO_{Z_A}$} determined by the Hilbert-Poincar\'e polynomial $P_A(t;\!\!\!k,\vec r\,)$ of the subscheme $Z_A\subset\FC^m$, which specifies how the D$1$-branes and the defects $\FH_{A^\circ}$ lie inside the D$(2m{+}1)_A$-branes. 
It follows that the instanton moduli space is isomorphic to the Quot scheme
\begin{align}
\frM_{k,\vec r\,}(\FC^m) \, \simeq \, \Quot^{P_A(t;\!\!k,\vec r\,)}_{\FC_A^m}(\CO^{\oplus r_A}) \ ,
\end{align}
independently of the choice of label $A\in\mathscr{C}_m$ since the noncommutative instanton equations are symmetric under permutation of the entries of $\vec r$.

The equivariant gauge theory on the Calabi-Yau $4$-fold $\FC^4$ is again defined by introducing the $\Omega$-background. The global symmetry group of the theory is then rotated to the maximal torus $\sT_{\vec t}=\big(\timesbig_{A\in\mathscr{C}_m}\,\sT_{\vec \tta _{A}}\big)\times \sT_{\vec \varepsilon}$ where $\vec \varepsilon =(\varepsilon_1,\dots,\varepsilon_4)$ and $\vec \tta_A=(\tta_{A,1},\dots, \tta_{A, r_A})$ are vacuum expectation values of the complex Higgs field $\Phi=\bigoplus_{A\in\mathscr{C}_m}\,\Phi_A$. Via equivariant localization the instanton partition function localizes onto the \smash{$\sT_{\vec t\ }$}-fixed points of the instanton moduli space $\frM_{k,\vec r}(\FC^m)$. A $\sT_{\vec t\ }$-invariant spiked/tetrahedron instanton of charge $k$ corresponds to a sum \smash{$\bigoplus_{A\in\mathscr{C}_m}\,\CI_A = \bigoplus_{A\in\mathscr{C}_m}\,\bigoplus_{l=1}^{r_A}\, \CI_{A,l}$} where  $\CI_{A,l}$ are  monomial ideals of codimension $k_{A,l}$ in the polynomial ring $\FC[w_1,\dots,w_4]$ labelled by the Coulomb parameters $\tta_{A,l}$, with \smash{$\sum_{A\in\mathscr{C}_m}\, k_A=\sum_{A\in\mathscr{C}_m}\,\sum_{l=1}^{r_A}\,k_{A,l}=k$}. They are in correspondence with arrays \smash{$\underline{\vec{\mathscr{Y}}}= (\vec {\mathscr{Y}}_{A})_{A\in\CCC_m}$} where  $\vec {\mathscr{Y}}_{A}=( \mathscr{Y}_{A,1},\dots,  \mathscr{Y}_{A,r_A})$ are $m$-dimensional coloured Young diagrams with $k_A$ boxes.

Consequently, the normalized character  associated to a sum of ideals $\CI_A$ is 
\begin{align}\label{eq:character_spiked}
\!\!\!\!\Ch_{\CI_A}(t) :\!&= \textstyle\prod\limits_{a\in A}\,(1-\ee^{\,t\,\varepsilon_a}) \ \Tr_{\mathscr{H}_{\CI_A}}\,\ee^{\,t\,\Phi_A} \nonumber \\[4pt]
&= \textstyle\sum\limits_{l=1}^{r_A} \, \ee^{\,t\, \tta_{A,l}} \, \Big(1-\textstyle\prod\limits_{a\in A}\,(1-\ee^{\,t\,\varepsilon_a}) \\ & \hspace{1cm}\times\textstyle\sum\limits_{(b_a)_{a\in A}	\in\mathscr{Y}_{A,l}}\,\ee^{\,t\,\sum\limits_{a\in A}\,\varepsilon_a\,(b_a-1)}\Big) \ . \nonumber
\end{align}
Instanton partition functions are calculated starting from the character \eqref{eq:character_spiked} in~\cite{Nekrasov:2015wsu,Nekrasov:2016ydq,Pomoni:2021hkn}.

\subsection{Instantons on Toric Varieties}\label{subsec:toric}

The equivariant gauge theory can be defined on any toric $n$-fold $M_n$ with isometries. There are two distinct classes, corresponding to compact and non-compact $M_n$. The noncommutative gauge theory is only understood so far in detail for the non-compact case of a toric Calabi-Yau $n$-fold. We sketch the main new features of the instanton partition functions in these cases, which now also capture the geometry of $M_n$ and extra characteristic classes of the noncommutative instantons.

A smooth toric Calabi-Yau $n$-fold $M_n$ is completely determined by its toric diagram $\Delta_{M_n}$, from which one may combinatorially construct $M_n$ from a set of rules which glue together $\FC^n$ patches. 
The set of $n$-valent vertices $V_{M_n}$ of ${\Delta_{M_n}}$ correspond to the $\sT_{\vec\varepsilon\,}$-fixed points of $M_n$, while the set of edges $E_{{M_n}}$ represent  $\sT_{\vec\varepsilon\,}$-invariant projective lines $\PP^1$ connecting a pair of fixed points with normal bundle $\bigoplus_{i=1}^{n-1}\,\CO_{\PP^1}(-m_i)$. The Calabi-Yau condition imposes $\sum_{i=1}^{n-1}\,m_i=2$. For each vertex $v\in V_{{M_n}}$ there is a local coordinate chart $\FC^n$ centered at the fixed point on which the torus $\sT_{\vec\varepsilon\,}$ acts with weights $\vec\varepsilon\,^v$. 

The idea is to solve the noncommutative instanton equations in each patch and then glue the local solutions together into a global contribution to the path integral. The instanton partition function that we describe below enumerates bound states of D$0$--D$2$--D$(2n)$-branes in $M_n$. Recall from Section~\ref{subsec:CohFT} that for $n=2,3$ this is tantamount to restricting to solutions of the noncommutative DUY equations (or assuming that $M_n$ contains only compact $2$-cycles). Geometrically, the formalism below describes the `equivariant vertex' for the `crystal melting' partition function, which enumerates subschemes of dimensions $0$ and $1$ in~$M_n$.

In this set up, $\sT_{\vec t\ }$-invariant instantons of $\sU(r)$ noncommutative gauge theory on each $\FC^n$ patch in the Coulomb phase are described by sums of monomial ideals \smash{$\CI_v=\bigoplus_{l=1}^r\,\CI_{v,l}$ in $\CO_{\FC^n}$} associated to infinite coloured $n$-dimensional Young diagrams $\vec{\mathscr{Y}}_v$, for $v\in V_{{M_n}}$ (with a suitably regularized hypercube count $\vert\vec{\mathscr{Y}}_v\vert$). They correspond to torsion free sheaves on $M_n$ supported on the subscheme $Z$ formed by $V_{{M_n}}$ and $ E_{{M_n}}$. 

Their contribution to the partition function is specified by the normalized version of the character $\Tr_{\CH_{\CI_v}}\,\ee^{\,t\,\Phi}$, which now depends on the asymptotic boundary conditions needed to glue the patches together. The boundary conditions are fixed by the asymptotics of $\vec{\mathscr{Y}}_v$ in the coordinate directions of $\FC^n$ labelling the corresponding edges emanating from the vertex $v$, which are expressed in terms of coloured $(n{-}1)$-dimensional Young diagrams $\vec\CCX_{a}$ for $a=1,\dots,n$. 
The character associated to the ideal $\CI_{v}$ which generalizes \eqref{eq:character} is then given by
\begin{align}\label{eq:characterv}
\begin{split}
\!\!\!\! \Ch_{\CI_v}(t)=&\textstyle\sum\limits_{l=1}^r\,\ee^{\,t\,\tta_l}\,\Big(1-\textstyle\sum\limits_{a=1}^n \, \textstyle\prod\limits_{\stackrel{\scriptstyle a'=1}{\scriptstyle a'\neq a}}^n\, \big(1-\ee^{t\,\varepsilon_{a'}^v}\big) \\
& \hspace{1.4cm} \times \textstyle\sum\limits_{\vec b'\in\CCX_{a',l}}\ee^{\,t\sum\limits_{a'\neq a}\varepsilon_{a'}^v\,(b_{a'}'-1)} \\ & -\textstyle\prod\limits_{a=1}^n\,\big(1-\ee^{\,t\,\varepsilon^v_a}\big)\,\textstyle\sum\limits_{\vec b\in\CCY_{v,l}}\,\ee^{\,t\,\vec\varepsilon\,^v\cdot\,\vec b}
\Big)  \ .
\end{split}
\end{align}

For each patch labelled by a vertex $v$, the calculation follows the steps described in Section~\ref{sec:partfns} for the computation of fluctuation determinants and the weighting by the gauge coupling $q$ of the top equivariant Chern class $\ch^{\sT_{\vec t\, ^v}}_n(\CE_{\CI_v})$, which gives the D$0$-brane charges. Each vertex contribution  has a factor associated with the vertex itself, as well as edge factors associated with the asymptotic Young diagrams. The edge factors are treated using both contributions associated to an edge coming from its two adjacent vertices. This gives the D$2$-brane charges through the equivariant Chern class \smash{$\ch^{\sT_{\vec t\, ^v}}_{n-1}(\CE_{\CI_v})$} which is computed from the coefficient of $t^{n-1}$ in the series expansion of \eqref{eq:characterv} near $t=0$. 

The compact $2$-cycles wrapped by the D$2$-branes are elements of $H_2(M_n)$, which is generated by lines $\PP^1$ labelled by the edges $e\in E_{M_n}$.
By Poincar\'e duality, they may be identified with the Pontryagin fluxes for the Chan-Paton gauge fields valued in the compactly supported cohomology $H_{\rm c}^{2n-2}(M_n,\RZ)$ in every Cartan direction $l=1,\dots,r$. These are determined (up to orientation) by the sizes of the boundary Young diagrams $\vert\CCX_{e,l}\vert$, which are counted by parameters $Q_e$ for each edge $e$ determined by the K\"ahler classes of $M_n$. In the following we denote the collection of these weights by $\underline{Q}:=(Q_e)_{e\in E_{M_n}}$.

\noindent\fbox{\small{$\boldsymbol{n=2}$}} \ 
On a toric surface $M_2$ of $\sSU(2)$ holonomy, the crystal and instanton partition functions are related but not quite the same~\cite{Cirafici:2009ga}.
The asymptotic profile of an ordinary infinite Young diagram $\CCY$ is simply a pair of non-negative integers $(d_1,d_2)$, which we extend to include arbitrary orientations corresponding to the magnetic charges of D$2$-branes wrapping spheres $\PP^1$. We denote a collection of magnetic fluxes through the edges of $M_2$ by $\underline{d}:=(d^e)_{e\in E_{M_2}}\in H^2_{\rm c}(M_2,\RZ)$, and the charges of a D$2$-brane labelled by $e$ inside the $r$ D$4$-branes as $\vec d\,^e:=(d^e_{1},\dots,d^e_{r})$.

The Nekrasov partition function of the pure $\CN=2$  $\sU(r)$ gauge theory on $M_2$ is then given by the ``master formula''~\cite{Nekrasov:2003vi,Gasparim:2009sns}
\begin{align}\label{eq:masterformula}
& \! Z_{M_2}^r(\Lambda,\underline{Q}\,;\!\!\vec\tta, \vec\varepsilon\,)^{\rm pure}  \\
&  = \textstyle\sum\limits_{{\underline{d}}_1,\dots,\underline{d}_r\in H^2_{\rm c}(M_2,\RZ)} \ \prod\limits_{e\in E_{M_2}}\,{Q}_e^{\,\sum\limits_{l=1}^r\,d^e_l} \, \Lambda^{\frac12\,\sum\limits_{l=1}^r\,\underline{d}_l\cdot C^{-1}\,\underline{d}_l} \nonumber \\
& \hspace{1cm} \times\!\!\textstyle\prod\limits_{v\in V_{M_2}} \, Z_{\FC^2}^r\Big(\Lambda;\! \vec{\tta} + \textstyle\sum\limits_{e\in E_{M_n}}\, w_e^v\,\vec d\,^e , \vec\varepsilon\,^v\Big)^{\rm pure} \nonumber
\end{align}
where $C$ is the intersection matrix of the compact divisors $D_e\in H_2(M_2)$, and $w_e^v$ are the weights of the $\sT_{\vec\varepsilon\,}$-equivariant line bundles $\CO_{M_2}(D_e)$ at the fixed point labelled by $v$.

The analogue expression in $\sU(r)$ Vafa-Witten theory on $M_2$ simplifies because of the $\vec t\,$-independence of the $\CN=4$ partition function on $\FC^2$, giving the factorization~\cite{Fujii:2005dk,Fucito:2006kn,Griguolo:2006kp,Bonelli:2007hb,Bruzzo:2009uc}
\begin{align}\label{eq:VWM2}
Z^r_{M_2}(q,\underline{Q}\,) = \big(\hat\eta(q)^{-\chi(M_2)} \, \Theta_{M_2}(q,\underline{Q})\big)^r \ ,
\end{align}
where $\chi(M_2)$ is the Euler characteristic of the toric surface $M_2$ and $\Theta_{M_2}(q,\underline{Q})$ is the theta-function defined by the second line of \eqref{eq:masterformula} (with the usual replacement $\Lambda\to q$). For the minimal ALE resolution of the toric $A_{p-1}$ orbifold singularity discussed in Section~\ref{subsec:orbifold}, the first factor in \eqref{eq:VWM2} is the contribution from regular instantons which are free to move throughout $M_2$; they have vanishing first Chern class and transform in the regular representation of the orbifold group $\sGamma=\RZ_p$.  On the other hand, the second factor is the contribution from the fractional instantons which represent anti-self-dual connections with non-trivial magnetic fluxes through the compact divisors $D_e$.

\noindent\fbox{\small{$\boldsymbol{n=3}$}} \ 
On a toric Calabi-Yau $3$-fold $M_3$ the gauge theory calculation agrees with the equivariant vertex/crystal calculation in Donaldson-Thomas theory~\cite{Maulik:2003rzb}. The asymptotic profile of a $3$-dimensional Young diagram $\CCY$ is a triple of ordinary Young diagrams $(\CCX_1,\CCX_2,\CCX_3)$. 

The instanton partition function for the $\sU(r)$ cohomological gauge theory on $M_3$ in the Coulomb phase is given by~\cite{Iqbal:2003ds,Cirafici:2008sn,Szabo:2011mj,Cirafici:2018jor}
\begin{align}\label{eq:DTM3}
\begin{split}
Z_{M_3}^r(q,\underline{Q}) &= \textstyle\sum\limits_{\underline{\vec\CCX}} \ \prod\limits_{v\in V_{M_3}} \, {\tt V}_{v,\underline{\vec\CCX}}(q) \\
& \hspace{1.5cm} \times \textstyle\prod\limits_{e\in E_{M_3}} \, {\tt E}_{\vec\CCX_e}(q,Q_e) \ ,
\end{split}
\end{align}
with the vertex contributions
\begin{align}
{\tt V}_{v,{\underline{\vec\CCX}}}(q) = \textstyle\sum\limits_{\vec \CCY_v \, \vert \, \partial\vec\CCY_v=\underline{\vec\CCX}} \, \big((-1)^r\,q\big)^{\sum\limits_{l=1}^r\,\vert\CCY_{v,l}\vert}
\end{align}
and the edge contributions
\begin{align}
\begin{split}
{\tt E}_{\vec \CCX_e}(q,Q_e) &= \big((-1)^r\,q\big)^{\sum\limits_{l=1}^r\,f_{\CCX_{e,l}}} \ Q_e^{\sum\limits_{l=1}^r\,\vert\CCX_{e,l}\vert} \\
& \quad \, \times (-1)^{\sum\limits_{l,l'=1}^r\, m_{1,e}\, \vert\CCX_{e,l}\vert\,\vert\CCX_{e,l'}\vert} \ ,
\end{split}
\end{align}
where $f_{\CCX_{e,l}}$ is an integer determined by the normal bundle over the compact $2$-cycle labelled by $e$.

The factorization of the Donaldson-Thomas partition function \eqref{eq:DTM3} is analogous to the splitting of the Vafa-Witten partition function \eqref{eq:VWM2}. 
This partition function (or more precisely its K-theory version on $M_3\times{\rm S}^1$) was extended in~\cite{Nekrasov:2014nea,DelZotto:2021gzy} to the general $\sU(r)$ gauge theory and in~\cite{DelZotto:2021gzy} to include D$4$-branes wrapping compact $4$-cycles of $M_3$, corresponding to the faces of the toric diagram~$\Delta_{M_3}$. 

\noindent\fbox{\small{$\boldsymbol{n=4}$}} \ 
The equivariant gauge theory approach to noncommutative instantons on a general toric Calabi-Yau $4$-fold $M_4$ has not been studied in detail yet. An equivariant vertex formalism is developed in~\cite{Cao:2017swr,Cao:2019tvv,Monavari:2022rtf} using geometric methods which enumerates D$0$--D$2$--D$8$-brane bound states in rank~$1$ Donaldson-Thomas theory on $M_4$.

\subsection{Further Developments}

There is a more extensive body of literature on noncommutative instantons in other settings which do not directly fit into the main line of development discussed in this paper, but which are certainly worthy of mention here. 

A close avatars of our theories in $4$ dimensions are the instantons on toric noncommutative manifolds such as spheres~\cite{Landi:2006qw,Brain:2012cj}, which uses techniques from the theory of isospectral deformations and noncommutative geometry in braided monoidal categories. By combining these approaches with techniques from toric geometry and noncommutative algebraic geometry, one can quantize toric varieties directly, which offers an alternative approach to noncommutative instantons compared to the patching construction discussed in Section~\ref{subsec:toric}~\cite{Cirio:2010ip,Cirio:2011eh,Cirio:2012sw}.

These types of noncommutative instantons come from $q$-deformations, of the type originally discussed in~\cite{Kapustin:2000ek}. They also appear in gauge theories on various noncommutative spaces based on quantum groups~\cite{Dabrowski:2000nz,Bonechi:2000ia,Frenkel:2002pm,DAndrea:2013kye}. Some of these theories have the curious feature of producing noncommutative moduli spaces of noncommutative instantons~\cite{Fiore:2006kh,Landi:2007ry,Brain:2009it}.

Finally, we mention for completeness that instantons have also been studied in gauge theories on fuzzy spaces and in matrix models~\cite{Balachandran:1999hx,Grosse:2001ss,Valtancoli:2006dw}.

\bmhead{Acknowledgments}

{\sc R.J.S.} thanks the editors Konstantinos Anagnostopoulos, Peter Schupp and George Zoupanos for the invitation to contribute to this special issue.
The work of {\sc R.J.S.} was supported by
the STFC Consolidated Grant ST/P000363/1. The work of {\sc M.T.} was supported by an EPSRC
Doctoral Training Partnership grant. 

\bibliography{NC-bibliography}


\end{document}